 \documentclass[sigconf, 9pt]{acmart}

\usepackage{subcaption}
\usepackage{enumitem}
\usepackage{amsfonts,amsmath,amsthm}
\usepackage{xurl}

\AtBeginDocument{%
  }

\newcommand{\xref}[1]{\S\ref{#1}}

\copyrightyear{2026}
\acmYear{2026}
\setcopyright{cc}
\setcctype{by}
\acmConference[MobiSys '26]{The 24th Annual International Conference on Mobile Systems, Applications and Services}{June 21--25, 2026}{Cambridge, United Kingdom}
\acmBooktitle{The 24th Annual International Conference on Mobile Systems, Applications and Services (MobiSys '26), June 21--25, 2026, Cambridge, United Kingdom}
\acmDOI{10.1145/3745756.3809210}
\acmISBN{979-8-4007-2027-7/2026/06}

\newcommand{\squishlist}{\begin{itemize}[itemsep=1pt,parsep=2pt,topsep=3pt,partopsep=0pt,leftmargin=0em, itemindent=1em,labelwidth=1em,labelsep=0.5em]}
\newcommand{\squishend}{\end{itemize}}

\begin{document}



\title{Fine-grained  Soundscape Control for Augmented Hearing}

\author{Seunghyun Oh,$^1$ Malek Itani,$^{1,2}$  Aseem Gauri,$^1$ Shyamnath Gollakota$^{1,2}$}
\affiliation{
\institution{$^1$Paul G. Allen School of Computer Science and Engineering, University of  Washington}
\institution{$^2$Hearvana AI}
\country{}
}




\begin{teaserfigure}
\centering
\vskip -0.1in
  \includegraphics[width=0.642\textwidth]{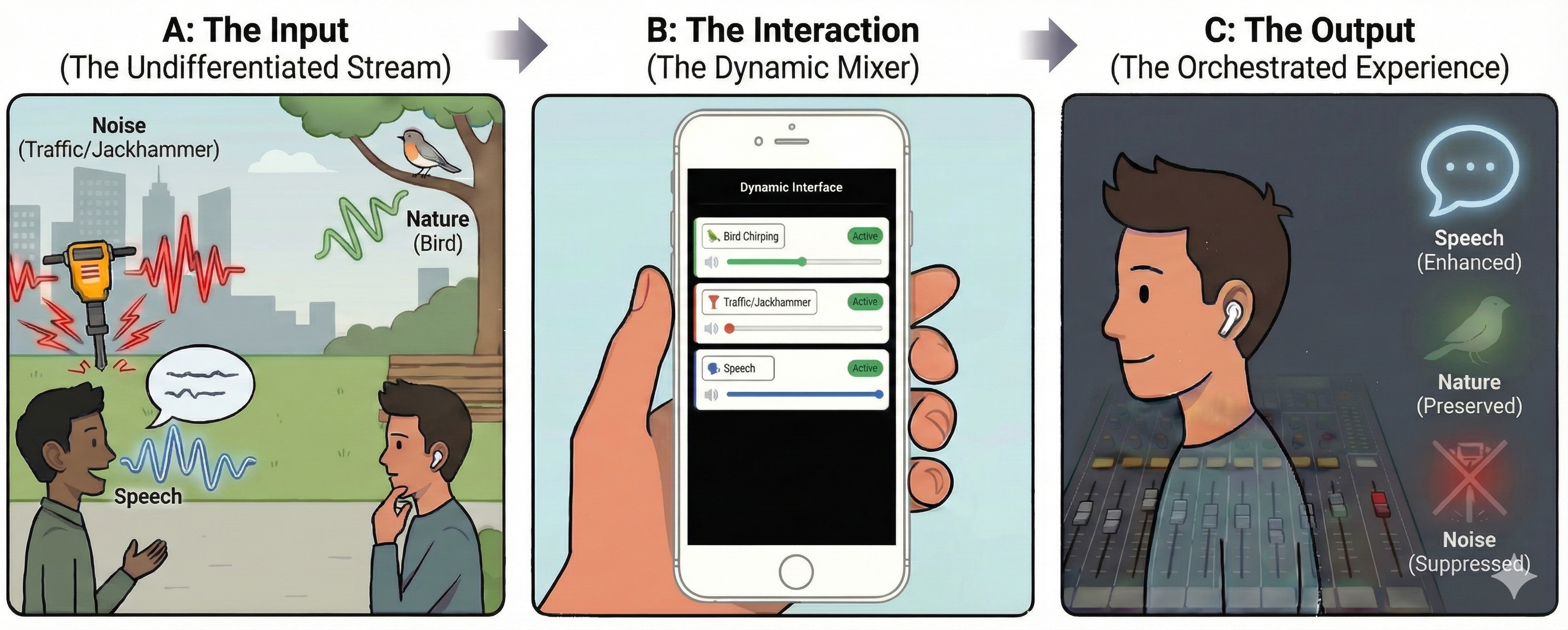}
  \vskip -0.1in
  \caption{Aurchestra transforms the auditory world into a programmable studio. \textmd{Unlike traditional hearables that offer binary noise cancellation (all-or-nothing), Aurchestra enables fine-grained soundscape control. (A) In a complex acoustic scene, (B) the system automatically detects active sound classes (e.g., speech, traffic, birds) and populates a dynamic interface. The user can then ``mix'' their reality in real-time, (C) independently suppressing interference (traffic) while increasing the volume of some  targets (speech) and maintaining others (nature), effectively acting as the audio engineer of their own life.}}
  \label{fig:teaser}
\end{teaserfigure}

\begin{abstract}

Hearables are becoming ubiquitous, yet their sound controls remain blunt: users can either enable global noise suppression or focus on a single target sound. Real-world acoustic scenes, however, contain many simultaneous sources that users may want to adjust independently. We  introduce Aurchestra, the first system to provide fine-grained, real-time soundscape control on resource-constrained hearables. Our system has two key components: (1) a dynamic interface that surfaces only active sound classes and (2) a real-time, on-device multi-output extraction network that generates separate streams for each selected class, achieving robust performance for upto 5 overlapping target sounds and letting users mix their environment by  customizing per-class volumes, much like an audio engineer mixes tracks. We optimize the model architecture for multiple compute-limited platforms and demonstrate real-time performance on 6~ms streaming audio chunks. Across real-world environments in previously unseen indoor and outdoor scenarios, our system enables expressive per-class sound control and achieves substantial improvements in target-class enhancement and interference suppression. Our results show that the world need not be heard as a single, undifferentiated stream: with Aurchestra, the soundscape becomes truly  programmable.

\end{abstract}







\maketitle

\section{Introduction}

Our acoustic environments are rich, dynamic, and often overwhelming~\cite{Vianna2015NoisePollution}. At any moment, a listener may want to tune in to a nearby sound, amplify an important cue such as an approaching vehicle, dampen distracting chatter, or simply enjoy the surrounding ambience. Yet today’s hearables, both commercial devices and research prototypes, offer only blunt controls: global noise-cancellation modes or a single target-sound focus~\cite{lookoncetohear,semantichearing}. In practice, this means users can either pick one sound or suppress all sounds. But the real world is not a toggle switch; {\it it is an orchestra of sounds.}


In this paper, we ask an intriguing question: what if users could mix and shape the sounds around them the way an audio engineer mixes tracks in a studio? Instead of hearing the world as one undifferentiated stream, imagine independently controlling the volume of speech, traffic, birds, music, alarms, and dozens of other sources, in real time, on a hearable device. Such expressive soundscape control would transform hearables from one-dimensional filters into tools that let users actively sculpt their auditory environment.

To make it easier for the user, the interface for expressive soundscape control should also be dynamic as the world itself. Rather than a static long set of controls, it must adapt as the user’s needs change throughout the day. A listener might begin in a café, enjoying ambient music while suppressing chatter; walk through a park where music is irrelevant but car honks must remain audible for safety; and then return home, where traffic no longer matters but door knocks and household speech do, while the vacuum cleaner does not. An ideal hearing interface should reflect these dynamics automatically, showing only the most relevant sounds, while still giving users full control over their preferences.

Prior work falls short of this vision. The closest system, Semantic Hearing~\cite{semantichearing}, enhances only one sound class at a time, relies on static category lists, and cannot support multiple classes or per-class volume control.   As a result, existing  systems provide only binary, all-or-nothing control, preventing users from fully  shaping their soundscape. Enabling true orchestration of the auditory world requires a fundamentally different  system design.

In this paper, we explore three key research questions:
\begin{enumerate}

\item Can a real-time extraction neural network output multiple target streams, one per user-selected class, within strict latency and power budgets?

\item What are the tradeoffs for such networks as a function of the compute capabilities of  diverse and rapidly evolving hardware platforms?

\item How can we automatically surface only the sound classes active in the current environment, reducing user effort and enabling dynamic, per-class control?

\end{enumerate}

We present {\it Aurchestra,} the first augmented hearing system to provide fine-grained, per-class soundscape control for hearables. Our technical contributions are:

\squishlist

\item {\bf Real-time multi-output target sound extraction.} Once users select the sound classes they want to hear, Aurchestra must extract each target in real time on sub-10 ms audio chunks. Prior sound extraction networks use large, attention-heavy architectures suited to smartphones rather than low-power hearables, and current real-time extractors \cite{waveformer,semantichearing,wakayama25_interspeech} produce only a single output stream, preventing per-class volume control. We address these challenges with two contributions. First, we replace attention with a dual-path time-frequency model conditioned on a multi-hot encoding of user-selected classes. Although dual-path architectures are common in speech separation~\cite{lookoncetohear,tfmlpnet,neuralaids}, they are rarely applied to environmental audio; our results show they outperform prior real-time approaches across diverse sound classes. Second, to enable independent gain control, we output one stream per selected class. Rather than producing outputs for all trained classes (e.g., 20+), which is inefficient, we limit the network to a small set of output streams  (e.g., 5) and map sound classes to streams using the ordering in the multi-hot encoding. We show that the model reliably learns this dynamic  mapping and outperforms the strategy of always outputting 20 streams, one for each trained sound class.

\item {\bf Model optimizations for diverse hardware platforms. } 
Aurchestra must run on diverse, rapidly evolving hardware platforms with varying compute capabilities. To address this,
we explore the neural architecture design space and develop hardware-tailored variants of our real-time extraction network. We evaluate architectures combining bidirectional LSTMs, MLP-Mixers~\cite{mlpmixer}, and dual-path modules, each offering different accuracy–latency tradeoffs depending on the hardware. Variants are profiled on an Orange Pi 5B and a Raspberry Pi 4B, which integrate with over-the-ear headphones, and the GreenWaves GAP9 AI accelerator in NeuralAids~\cite{neuralaids}. For each platform, we select and optimize a model capable of processing 6~ms audio chunks in real-time.

\item {\bf Dynamic interface for augmented hearing control.}  Finally, Aurchestra uses a  sound event detection model that periodically identifies the sound classes present in the environment. Rather than forcing users to navigate long static lists~\cite{semantichearing}, it surfaces only the active classes on the companion device (e.g. phone), reducing effort and cognitive load. Users can then tap the classes they care about and adjust each one's volume independently for fine-grained control. A key challenge is reliably identifying sound classes in dense, overlapping scenes: existing classifiers, trained mostly on isolated or lightly mixed sounds, degrade sharply when multiple sources co-occur (see~\xref{sec:SED:comp}). We address this by fine-tuning state-of-the-art transformer models on heavily overlapped mixtures, enabling robust multi-class, multi-instance detection and improving accuracy from 63.8-81.5\% to 93.2\% on scenes with five simultaneous   overlapping target sounds.

\squishend

{\bf Key findings.} We train our models on  20  sound classes, including sirens, baby cries, speech, vacuum cleaners, alarm clocks, and bird chirps and evaluate Aurchestra in real-world indoor and outdoor environments across multiple user studies. Our results are as follows.

\squishlist
\item Aurchestra outperforms the prior real-time single-target baseline, achieving superior signal quality (11.99 dB vs 7.29 dB SNRi) while using less than half the parameters (0.5M vs 1.2M). Furthermore, our system maintains stable performance when extracting upto 5  simultaneous output streams, validating its ability to let users mix multiple distinct sound classes in real time.
\item  Our hardware-optimized networks process 6~ms audio chunks in a streaming manner and achieve inference times of 5.22, 4.47, and 5.23~ms on Orange Pi, Raspberry Pi, and GAP9, respectively. On NeuralAids, the model consumes  56~mW, demonstrating that Aurchestra enables efficient real-time soundscape control even on low-power hearables.

\item We evaluate the system in-the-wild with participants wearing headsets moving in previously unseen indoor and outdoor scenarios. Subjective listening studies with participants ($n=17$)  show that Aurchestra yields substantial improvements in background-noise suppression (+1.54 points) and overall listening experience (+0.95 points)  compared to the baseline, without introducing noticeable distortion.
\item  Another user study ($n=7$) evaluating our dynamic interface running on a smartphone in real-time   demonstrates that it significantly lowers interaction overhead. By automatically detecting and surfacing only active sound classes, the interface reduces the time required for users to select target sounds by 67.9\% compared to a static interface.

\squishend

Looking ahead, we envision augmented hearing systems that not only separate and remix the world in real time, but also learn user preferences, anticipate intent, and integrate seamlessly into everyday acoustic life. Aurchestra takes an important first step toward this broader vision.


\section{Related Work}

Our paper is related  to prior work on source separation~\cite{waveformer,music1,music2,itani25_interspeech}, accessibility~\cite{10.1145/3643834.3661556,10.1145/3706598.3714268}, and hearables~\cite{lookoncetohear,soundbubble}. Here we discuss the closest technical  works to our research.

\vskip 0.03in\noindent{\bf Target sound extraction.} Recent deep-learning methods leverage cues from audio~\cite{delcroix2022soundbeam,gfeller2021one}, text~\cite{kilgour2022text,liu2022separate}, images~\cite{gao2019co,xu2019recursive}, onomatopoeia~\cite{okamoto2022environmental}, one-hot labels~\cite{2020arXiv200605712O}, and semantic or spatial embeddings~\cite{chen25l_interspeech}. However, these systems operate offline on full audio clips ($\ge$1 s), making them incompatible with the stringent low-latency streaming requirements of real-time hearable devices. 

More recent efforts explore generative models~\cite{diffuse1,diffuse2,10890129}, but these approaches are computationally heavy for our target hearable platforms. Similarly, foundational audio models such as AudioLM~\cite{audiolm}, UniAudio~\cite{uniaudio}, and AudioFlamingo~\cite{goel2025audioflamingo3advancing} support broad audio tasks including continuation, generation, editing, and audio-level reasoning across speech, music, and environmental sounds. While powerful, they are large (100M–8B parameters), exceed the compute limits of hearable devices, and cannot satisfy the sub-20 ms streaming latency required for hearing applications.

The closest prior work is Semantic Hearing~\cite{semantichearing}, which shows that binaural target-sound extraction can run on smartphones. However, it does not support fine-grained soundscape control. Our work differs in four key ways: 
1) \cite{semantichearing} supports only a single target class in the environment, whereas we support multiple sound classes simultaneously. 2) It uses an all-or-nothing control framework; in contrast, we produce separate output streams per class, enabling independent fine-grained control for each target class.
 3) It requires users to manually choose one class from a static list; we introduce a dynamic  interface that detects classes present in the environment, reducing selection burden and improving usability.
 4) It is built on the Waveformer architecture~\cite{waveformer}, whose attention mechanism runs on smartphone GPUs but is difficult to deploy on the tiny AI accelerators used in hearing aids and earbuds~\cite{neuralaids}.

\vskip 0.03in\noindent{\bf Hearing systems for speech enhancement.} Prior work on hearables has primarily focused on improving speech quality in the presence of interfering speakers and noise. ClearBuds~\cite{chatterjee2022clearbuds} and NeuralAids~\cite{neuralaids} improve  speech quality in the presence of noise for teleconferencing and hearing aid applications, respectively. \cite{lookoncetohear,soundbubble,proactivehearing} focus on extracting target speakers in the presence of interfering speakers. All of these systems treat non-speech sounds as undifferentiated noise. In contrast, our work performs real-time semantic understanding of diverse  sounds and provides fine-grained control for shaping the user's soundscape.

\vskip 0.03in\noindent{\bf Hearable platforms and acoustic applications.} Prior work has developed platforms to support earable research~\cite{10.1145/3447993.3448624,10.1145/3372224.3419197,esense,10.1145/3544793.3563415,montanari2024omnibudssensoryearableplatform}. Other systems leverage sound for activity recognition in wearables and smart homes~\cite{bodyscope,soundsense,samosa,ubicoustics,dhruv,dhruv2,10.1145/3643834.3661556}, but they do not satisfy the low-latency streaming requirements of hearing applications. Research in our community has also explored in-ear sensing~\cite{earablegoogle,act1,chris1} as well as  a range of medical and health applications~\cite{10.1145/3498361.3538935, yang_liu_stuchbury-wass_ciliberto_röddiger_butkow_pullin_panariti_ma_mascolo_2025,10.1145/3636534.3649366,10.1145/3560905.3568084,infection,oae,oaebuds,pullin2025earecg}, which, while complementary to our work, highlight the versatility  of earables as a powerful  platform.


\section{Aurchestra}


We first describe our  real-time multi-target sound extraction model (\xref{sec:network}) and hardware-specific  optimizations (\xref{sec:hwsw}), followed by the training methodology (\xref{sec:datasets}) to generalize to previously unseen wearers and real-world environments.
Finally, we describe our dynamic interface design (\xref{sec:interface})

\subsection{Multi-Output Sound Extraction}\label{sec:network}
{
The goal here is to separate distinct audio classes into individual channels for independent manipulation, mixing, and playback. To achieve this, the system must meet two key requirements: (1) maintaining a low total latency  less than 20~ms between the acoustic environment and the audio playback to ensure the user does not perceive a difference, and (2) processing audio chunks in real-time, such that each chunk is fully processed before the subsequent chunk arrives.} 

\subsubsection{Problem Formulation} We are given a length-$T$ binaural mixture $x(t) \in \mathbf{R^{2\times T}}$ which has a known subset of $k\le \mathcal{K}$ target sound classes in the presence of interfering sounds, $n(t)$, from background sound classes. Here $\mathcal{K}$ is the total number of target sound classes known to the system. Let $s_i(t)\in \mathbf{R}^{2\times T},\ i = 1, \dots, k$ be  the  signal corresponding to the $i$-th class. The mixture signal $x(t)$ can then be written as, $x(t) = \sum_{i=1}^ks_i(t) + n(t)$.

Given a set of per-class volume modifiers $v \in \mathbf{R_{\ge0}}^k$, our goal is to compute a single-channel audio mixture $\hat{x}(t)$ comprised of only the $k$ signals mixed with the appropriate per-class volume modifiers and averaged left and right channels: $\hat{x}(t) = [0.5, \ 0.5]\sum_{i=1}^k v_i s_i(t)$. At inference time, incoming audio is processed in small chunk of  samples, so we can represent all audio signals as a concatenation of ${N}$ audio chunks, i.e., $x(t)=[x^1, \dots, x^{N}]$. Our goal is then to design a system $\mathcal{S}$ that produces output chunks of $\hat{x}(t)$ from input chunks of $x(t)$ given a multi-hot encoding of target classes $q\in\{0,1\}^\mathcal{K}$, such that $q_i=1$ if the $i$-th class is a target class, and per-class volume modifiers $v$: $\hat{x}^j = \mathcal{S}(x^j, q, v)$.

\subsubsection{Our  multi-output approach}

We address this problem by decomposing the audio mixture into $k$ individual streams, each representing a single sound class, and then forming the final estimate $\hat{m}(t)$ as a weighted sum of these streams. To enable this, we train a neural network $\mathcal{N}$ conditioned on the multi-hot encoding $q$.

A straightforward approach would be to have the network produce $\mathcal{K}$ output streams, one for each sound class in the training set. However, this is inefficient: if $\mathcal{K} = 20$, the model must always generate 20 streams, even though only a few classes are likely present in any given environment, leaving most outputs unused. Instead, motivated by the observation that real environments contain only a small subset of relevant target sounds, we design $\mathcal{N}$ to generate  $\mathcal{O}$ output streams, where $\mathcal{O} \ll \mathcal{K}$. Accordingly, $q$ is now constrained to select at most $k \le \mathcal{O}$ target classes.

This design offers two major advantages. First, it reduces computational overhead by requiring far fewer output streams. Second, it improves learning efficiency, especially for smaller models intended for compute-constrained devices,  since models with fewer output heads converge more reliably and perform better, as confirmed in our evaluation (see~\xref{sec:modelevals}).

\subsubsection{Mapping outputs to classes}

A smaller number of output streams requires mapping the target sounds in the environment to the correct output streams. When the number of output streams is equal to the total number of target classes in the training set, $\mathcal{K}$, we can constantly map the $i$-th sound class (e.g., in alphabetical order) to the $i$-th output channel. However, if the number of output streams $\mathcal{O}$ is smaller than $\mathcal{K}$, then mapping between sound classes and their corresponding output channel will change based on the other target classes in the multi-hot vector.

To resolve this, we train the network to dynamically assign a target sound class to the output stream corresponding to its alphabetical order among the other chosen sound classes. If there are fewer target classes than output streams ($k < \mathcal{O}$), only the first $k$ streams are used and the remaining streams are ignored for optimization purposes. For example, if $\mathcal{O}=3$, $k=2$, and the target classes are "\underline{c}at" and "\underline{d}og", the first output stream would correspond to the "cat" sound class, the second one to "dog" and the third is unused.
This approach has two key benefits to having a fixed, deterministic output stream assignment: 1) it removes the need to use Permutation Invariant Training techniques~\cite{neuralaids} which can lead to slower and unstable training, and 2) it avoids the need for a stitching algorithm to reorder channels after every successive inference during deployment.

\begin{figure}[t!]
\centering
\includegraphics[width=0.99\linewidth]{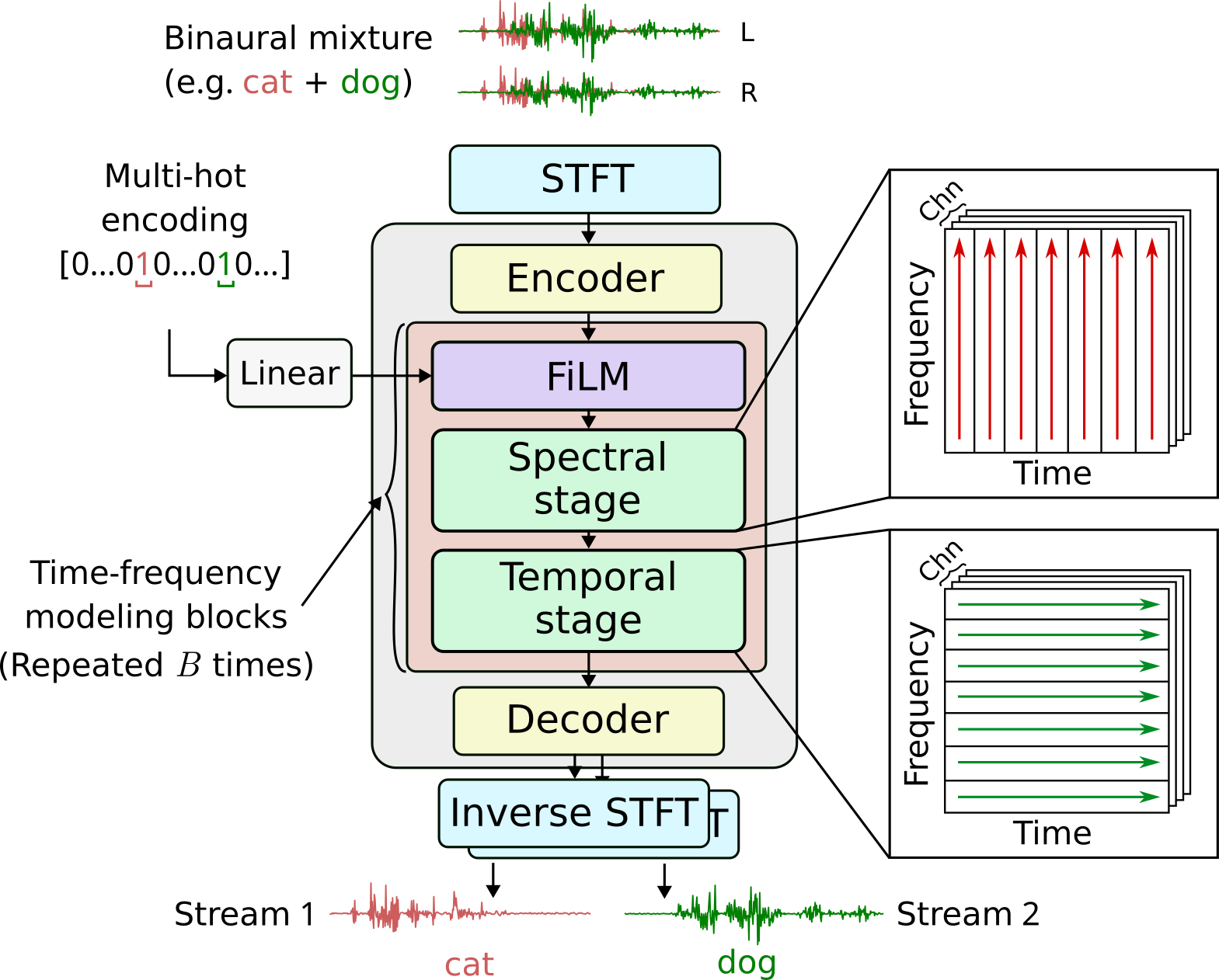}
\vskip -0.15in
\caption{\textmd{Multi-output sound extraction  architecture.}}
\label{fig:modelarch}
\vskip -0.2in
\end{figure}

\subsubsection{Network Architecture}\label{sec:arch}
The network in Fig.~\ref{fig:modelarch} processes incoming audio in the time-frequency (TF) domain. Incoming audio chunks $x^{j}$ are first transformed using a short-time frequency transform (STFT) into TF-representations $X^j$. The real and imaginary components are concatenated along the channel dimension, projected onto a latent space using a learned convolutional encoder $\mathcal{E}$, and successively processed using $B$ time-frequency modeling blocks. Each block consists of two stages: 1) a spectral stage which models the frequency sequences at every time frame, and 2) a temporal stage which models time sequences for every frequency bin. Finally, we use a transpose convolution decoder to generate the single-channel, time-frequency estimate $S^j_i$ for every selected target class $i$, and we use an inverse STFT (ISTFT) to obtain the time-domain output streams $s^j_i$.

To minimize the algorithmic latency, we utilize a dual-window STFT approach~\cite{lowlatse}. We process audio in chunks (hop length) of size $L_c=$6~ms, with $L_F=$4~ms of overlap with future chunks (lookahead) and $L_B=$6~ms of overlap with past chunks (lookback). Using a standard STFT with non-zero padded windows, the overlap-add step of the ISTFT would produce a total algorithm latency of $L_B + L_C + L_F=$16~ms. To reduce this, we make two key changes to construct the synthesis window: 1) the first $L_B$ samples are set to zero to prevent information flow between future chunks to the current chunk, and 2) the remaining $L_C + L_F$ samples are recomputed for perfect signal reconstruction following~\cite{neuralaids}. The resulting algorithmic latency is $L_C + L_F = $10~ms. 
   
    Beyond the $L_F$ lookahead samples, the neural networks do not utilize information from any additional future samples. This is done by using causal encoders and decoders with appropriate padding. Additionally, the temporal stage in all networks is a unidirectional LSTM (applied independently to all frequencies) followed by a linear projection from the hidden state $H$ back to the latent dimension $D$.
    
    We condition the model  on the multi-hot encoding using feature-wise linear modulation (FiLM)~\cite{film}. Specifically, we first use a linear layer to transform the multi-hot encoding into an embedding vector $Q \in R^D$. This vector is then used to condition the network to extract the selected target classes using FiLM. To effectively propagate conditioning information, we experiment with three different FiLM layer placements: 1) A single FiLM layer after the encoder, 2) one FiLM layer before every time-frequency modeling block, and 3) one FiLM layer before every time-frequency modeling block except the first (see ~\xref{sec:modelevals}).



\subsection{Hardware-Specific Model Optimizations}\label{sec:hwsw}

Next we explore the design space of the neural network architecture and components to design three different models that are customized to three different hardware platforms.

\vskip 0.03in\noindent{\bf Orange Pi model.} This model  is intended for deployment on an Orange Pi 5B, which uses an Arm Cortex-A76 CPU at 2.4~GHz and is our most powerful compute platform. To fully utilize its compute capabilities, the encoder $\mathcal{E}$ uses a 3$\times$3 causal convolution followed by layer normalization. The spectral stage consists of layer normalization, followed by a bidirectional LSTM with the same hidden dimension as the temporal stage to model the frequency sequence. A linear layer projects the activations back to the latent dimension. For this network, we use $D=32$, $H=64$, $B=6$. We use  strategies such as caching convolution buffers used in prior work~\cite{soundbubble,lookoncetohear,neuralaids} to further reduce runtime.


\vskip 0.03in\noindent{\bf Raspberry Pi model.} This network is intended for deployment on an Raspberry Pi 4B, which uses an Arm Cortex-A72 CPU at 1.8~GHz. We use a network similar to the Orange Pi model, with one major difference: we compress the number of frequency bins used five-fold in the spectral stage via a pair of strided convolution and transpose convolution layers. This reduces the number of frequency steps we need to process. Additionally, we set the network hyperparameters to $D=16$, $H=64$ and $B=3$.  For both Orange Pi and Raspberry Pi models, we deploy our networks using ONNXRuntime, which we found to be the fastest inference runtime.

\vskip 0.03in\noindent{\bf NeuralAids model.} This network is intended for deployment on the recent NeuralAids~\cite{neuralaids} platform which has AI accelerators for on-device streaming audio processing. It uses the GreenWaves GAP9, a dedicated low-power accelerator with a RISC-V compute cluster clocked up to 370~MHz. GAP9 has one notable feature: It has 10 cores for parallel processing, one of which is highly optimized for parallel 8- and 16-bit fixed-point operations. As a result, certain parallelizable layers can run much faster on GAP9 than they would on the previous two platforms. While the model in~\cite{neuralaids} uses a dual-path design similar to the Raspberry Pi model for GAP9, we replace the bidirectional LSTM with two repeated MLP-Mixer blocks~\cite{tfmlpnet} to better leverage the chip’s parallel processing capabilities.  In contrast to the sequential nature of LSTMs, MLP-Mixers consist of highly parallelizable linear layers applied alternately along the channel and frequency dimensions. Additionally, we implement the temporal processing stage batch LSTMs with Conv-Batched LSTMs~\cite{tfmlpnet} which achieve better parallelization with GreenWaves' model conversion tools. By exploiting parallelization, the resulting network runs much more efficiently on GAP9. Finally, we discard all layer normalization. We use the following hyperparameters: $D=32$, $H=32$, $B=6$.

\subsection{Training methodology}\label{sec:datasets}


\subsubsection{Sound Classes and Datasets}


We selected sound classes based on the AudioSet ontology~\cite{audioset}. Each sound class node has a unique AudioSet ID and may contain one or more child nodes representing more specific  classes.


{\bf Target sounds (20 classes).} We selected 20 sound classes to functionally cover everyday acoustic
contexts across representative scenarios. During urban commutes, users may
wish to preserve safety-critical sounds (siren, car horn) while
suppressing traffic noise and construction sounds (hammer, glass breaking).
At home, users may want to stay aware of conversational speech, baby cries,
and pet sounds (dog, cat) while filtering out appliance noise (typing,
toilet flush). In outdoor settings such as parks or beaches, users can
amplify nature sounds (birds chirping, ocean, cricket) while keeping
speech intelligible---illustrating that the system supports not only noise
suppression but active enhancement of desired sounds. Each class was mapped
to the semantically closest label in the AudioSet
ontology~\cite{audioset}, yielding 20 categories that humans can
distinguish with reasonably high accuracy: alarm clock, baby cry, birds
chirping, car horn, cat, rooster crow, typing, cricket, dog, door knock,
glass breaking, gunshot, hammer, music, ocean, singing, siren, speech,
thunderstorm, and toilet flush. A user preference survey (Appendix~A)
further corroborates this selection: participants' most commonly desired
sounds (speech, music, nature, safety cues) and unwanted sounds (traffic,
typing, babble) align closely with our target and interfering class sets.

{\bf Interfering sounds (141 classes).}
In practice, countless background sounds appear that fall outside the target categories. These interfering noises can stem from many different sources, making it unrealistic to list them exhaustively. To help the model handle such variability, we selected 141 interfering classes based on the AudioSet hierarchy. Viewing the AudioSet ontology as a directed acyclic graph, where edges connect each parent class to its children, we identified interfering classes as all nodes that share no path with any of the 20 target categories. This   prevents semantic overlap between interfering and target classes.


{\bf Datasets, preprocessing and data split.}
To obtain coverage for all 20 target classes and interfering categories, we drew from four  datasets.  FSD50K dataset~\cite{fsd50k} with more than 51,000 audio clips spanning 200 general-purpose sound categories. ESC-50~\cite{esc50} with 2,000 environmental audio samples grouped into 50 classes and arranged into five cross-validation folds. MUSDB18 \cite{musdb18} with 150  music tracks along with isolated streams for vocals and  instruments. Lastly, the DISCO dataset~\cite{disco} with  real-world noise recordings. 


Each class from every dataset was mapped to the most semantically appropriate AudioSet label whenever possible. For FSD50K, we further filtered out any recordings that consisted of mixtures of multiple sound sources so that only clean, single-source examples remained. For  MUSDB18, we separated each track into its vocal and instrumental stems and labeled them as “Singing’’ and “Melody’’. All audio was then divided into 15-second segments, and any segment falling below a power-based silence threshold was removed.

Each dataset was first split into training, validation, and test sets, then merged into a unified corpus. FSD50K and MUSDB18 used a 90:10 split within their development sets for training and validation, with the evaluation sets used for testing. ESC-50 employed folds 1–3 for training, fold 4 for validation, and fold 5 for testing. DISCO was divided 60\%/7\%/33\% into training, validation, and test sets. 


{\bf Binaural Data Synthesis.} We used the CIPIC dataset~\cite{CIPIC},  which provides head-related transfer functions (HRTFs) from 43 subjects. For each training example, we randomly chose one subject to capture anatomical diversity, then independently assigned each sound source a random direction from that subject’s available measurements, allowing multiple sources to share a direction. The corresponding left and right ear impulse responses were then used to convolve each mono signal, yielding the final binaural audio.

Training data were generated on-the-fly with Scaper~\cite{scaper}, producing 20,000 training, 2,000 validation, and 2,000 test samples. Each 5-second binaural mixture contained 1-5 target classes and 1–2 interfering classes. Target and interfering events lasted 3–5 seconds, were placed at random offsets, and were mixed with continuous urban background noise from TAU Urban Acoustic Scenes 2019 dataset, normalized to –50 LUFS. Silent segments were removed, targets were mixed at 5–15 dB SNR, and interferers at 0–10 dB SNR. Audio was resampled to 16 kHz.

Each source was spatialized by selecting a random CIPIC subject and direction, then convolving the mono audio with the  left/right HRTFs. The final mixture was obtained by summing all convolved sources with peak normalization. 

\subsubsection{Training hyperparameters and loss functions}
We train  our  models using  AdamW  optimizer for 200 epochs. The learning rate is initially set to 1e-3, and we halve it if the validation loss does not improve after 4 consecutive epochs. Our loss function  combines L1-loss and a multiresolution spectrogram loss with perceptual weighting, provided by~\cite{auraloss}. These are computed between the output streams of target classes and their corresponding clean ground truths.

\subsection{Dynamic Interface Design}\label{sec:interface}

The goal for our interface is to help users understand their acoustic environment and quickly choose what they want to hear. Such an interface should ideally translate raw acoustic complexity into a set of actionable, meaningful options, while staying responsive enough to be used. 

To do this, we design a dynamic interface that is continuously populated based on the real-world sounds surrounding the user. A phone app, paired with the hearable, acts as the orchestration hub. Incoming audio is streamed to the phone, where a lightweight Sound Event Detection (SED) model analyzes it in real time. Rather than presenting a static or predetermined list of categories, the interface surfaces only the sound classes that are currently active or recently detected. This design ensures that the user is not overwhelmed with irrelevant options, and instead receives a concise, context-aware menu of the sounds available for control.



\subsubsection{Sound event detection model}

For the interface to remain responsive, our sound event detection (SED) model must meet several  requirements. First, it must accurately identify active sound classes from short audio segments, so the interface can update promptly without relying on long recording windows. Second, it should maintain reliable performance in scenes containing multiple target events, where several sound sources may be present at once. Third, the model must remain robust when these sources overlap significantly for extended periods, as is common in real-world acoustic settings (e.g., a jackhammer running continuously with speech and traffic sounds  in the background). 

To meet these requirements, we build on the  Audio Spectrogram Transformer (AST)~\cite{AST}, which uses a Vision Transformer (ViT) backbone to model complex time–frequency patterns. In~\xref{sec:SED:comp}, we compare AST against alternative SED models such as YAMNet~\cite{yamnet2020}, which demonstrates inferior performance under our conditions. The original AST is pre-trained on AudioSet~\cite{audioset}, a large-scale multi-label dataset of YouTube audio clips, enabling it to detect multiple sound events within the same recording.

\xref{sec:SED:comp} reveals  key limits of existing models when deployed in our target scenario. While they perform well on isolated sound classes, their accuracy drops substantially as the number of concurrent events increases. In scenes featuring sustained overlap between multiple sound classes,  both precision and recall deteriorate further as the  window size shortens, precisely the opposite of what our interface requires.

\begin{table}[t!]
  \caption{Model comparison for target sound extraction. \textmd{Evaluation performed with 20 target classes, single source in mixture, and single output channel. }}
  \label{tab:model_comparison}
  \vskip -0.1in
  \centering
  \setlength{\tabcolsep}{6pt}
  {\footnotesize
  \begin{tabular}{ l c c c }
    \toprule
    \textbf{Model} & \textbf{\# Params (M)} & \textbf{SNRi (dB)} & \textbf{SI-SNRi (dB)} \\
    \midrule
    OrangePi & 0.498 & 11.99 $\pm$ 7.04 & 11.27 $\pm$ 8.21 \\
    Raspberry Pi & 0.208 & 10.13 $\pm$ 4.01 & 7.72 $\pm$ 5.17 \\
    NeuralAids & 0.502 & 9.75 $\pm$ 4.80 & 7.60 $\pm$ 6.48 \\
    Waveformer \cite{semantichearing} & 1.207 & 7.29 $\pm$ 6.11 & 5.58 $\pm$ 7.67 \\
    \bottomrule
  \end{tabular}
}
  \vskip -0.2in
\end{table}

\subsubsection{Fine-tuning procedure}\label{sec:finetuning}
We fine-tune the AST model on our domain-specific dataset to improve its robustness under short windows, multi-event mixtures, and long-duration overlaps. We use the dataset  in \xref{sec:datasets} where each training example is a 5-second binaural mixture created by combining  1-3 target sound classes and  1-2 interfering classes. Target and interfering categories were chosen independently from their respective pools: the 20 designated target categories and the 141 non-target categories.  Sound events drawn from both target and interfering classes lasted 3–5s and were inserted at random onset times within the 5s mixture.


We separated the AST model into an encoder and a classifier. The number of output nodes in the classifier was modified to match 20 predefined target classes. We applied different learning rates to the two modules: the encoder's learning rate was set lower than the classifier's learning rate to preserve the pre-trained feature representations while adapting to the new task. Training was conducted for 50 epochs using AdamW optimizer with an initial learning rate of 1e-4 for the classifier and 1e-6 for the encoder.

\subsubsection{Reducing perceived latency.} 
The time it takes for the interface to display the correct sound classes in the user’s environment depends on two components: the algorithmic latency $T_a$ and the computational latency $T_c$. The algorithmic latency is the duration of the audio chunk required by the SED model; the computational latency is the time needed to process that chunk on the companion device (e.g., a smartphone). The total latency experienced by the system is  $T_a+T_c$.

Unlike computational latency, algorithmic latency cannot be improved simply by using faster hardware. \xref{sec:sed:runtime} shows that  reliable precision and recall, with multiple sound classes overlap, requires  at least 3–5 seconds audio segments. This creates a  tension: longer segments improve accuracy but increase latency, which can make the interface feel sluggish.

To reduce the perceived latency for the user, we adopt a staggered buffering strategy. Instead of waiting for the current $T_a$-second segment to complete, we run the SED model on the previous audio chunk while the next chunk is still being recorded. The interface is then populated using the results from this preceding window. This pipelined approach removes the algorithmic latency from the user’s experience, making the interface feel responsive even though the model still operates on multi-second audio segments.

\section{Experiments and Results}

\begin{table}[t!]
  \caption{Aurchestra's performance across different mixture complexities and output channel configurations. \textmd{Total target classes: 20.} }
  \label{tab:orangepi_eval}
  \vskip -0.1in
  \centering
  {\footnotesize
  \setlength{\tabcolsep}{6pt}
\begin{tabular}{ c c c c }
    \toprule
    \textbf{\#Targets} & \textbf{\#Outputs} & \textbf{SNRi (dB)} & \textbf{SI-SNRi (dB)} \\
    \midrule
    1 & 1 & 11.99 $\pm$ 7.04 & 11.27 $\pm$ 8.21 \\
     & 5 & 9.56 $\pm$ 6.74 & 8.53 $\pm$ 7.85 \\
     & 20 & 3.51 $\pm$ 6.16 & 3.10 $\pm$ 6.23 \\
    \midrule
    2 & 1 & 13.10 $\pm$ 4.54 & 11.93 $\pm$ 5.77 \\
     & 5 & 11.81 $\pm$ 4.07 & 10.29 $\pm$ 6.00 \\
     & 20 & 6.63 $\pm$ 3.30 & 6.04 $\pm$ 3.41 \\
    \midrule
    3 & 1 & 12.83 $\pm$ 3.45 & 11.10 $\pm$ 4.85 \\
     & 5 & 12.38 $\pm$ 3.15 & 10.27 $\pm$ 4.62 \\
     & 20 & 8.36 $\pm$ 2.49 & 7.58 $\pm$ 2.60 \\
    \midrule
    4 & 1 & 12.58 $\pm$ 2.91 & 9.79 $\pm$ 4.72 \\
     & 5 & 12.64 $\pm$ 2.62 & 9.87 $\pm$ 4.18 \\
     & 20 & 9.10 $\pm$ 1.89 & 8.11 $\pm$ 2.04 \\
    \midrule
    5 & 1 & 12.61 $\pm$ 2.50 & 9.33 $\pm$ 4.53 \\
     & 5 & 13.04 $\pm$ 2.21 & 9.84 $\pm$ 3.93 \\
     & 20 & 10.03 $\pm$ 1.74 & 8.86 $\pm$ 2.05 \\
    \bottomrule
  \end{tabular}}
  \vskip -0.2in
\end{table}

\subsection{Benchmarking Target Sound Extraction}\label{sec:benchmarks}

\subsubsection{Evaluation Metrics} Our evaluation of the   extraction model includes both separation  and system performance. 

\squishlist
\item {\it SI-SNRi.}
Scale-Invariant Signal-to-Noise Ratio (SI-SNR) is a popular metric for source separation tasks and measures how closely the extracted audio signal matches the original clean signal, and SI-SNRi is the improvement in output signal quality relative to the input mixture. 

\item{\it SNRi.}
This is the improvement in the Signal-to-Noise Ratio between the target sound component and residual noise component in decibels (dB), comparing the output signal to the input mixture.  

\item{\it Number of parameters.}
This reflects the model’s complexity and capacity. More parameters may improve performance but increase memory use and latency. 

\item{\it Runtime.}
This is the inference time for a single audio chunk in milliseconds. For real-time streaming processing, the inference time must be shorter than the audio chunk duration. 

\squishend

\subsubsection{Model Comparison}

We compare four models for different hardware platforms, evaluated on a test set of 20 target sound classes with single-source mixtures and single-output channel configuration, following the setup of prior work~\cite{semantichearing} for fair comparison.

\squishlist
\item {\it Orange Pi model:} This model has  0.498M parameters and is  designed for the Orange Pi platform which offers the higher computational capacity.
\item {\it Raspberry Pi model:} A compressed variant of above model designed for the Raspberry Pi platform, featuring frequency-domain compression, reduced number of layers, and smaller latent dimensions to meet computational constraints.
\item {\it NeuralAids model:} Our custom model with 0.502M parameters, designed for the NeuralAids  platform.
\item {\it Waveformer (baseline):} The baseline model from Semantic Hearing~\cite{semantichearing} with 1.207M parameters. 
\squishend

Table~\ref{tab:model_comparison} shows the source separation performance for each model. The Orange Pi model achieves the best performance, outperforming both our lightweight NeuralAids model  and the Waveformer baseline. Notably, our Orange Pi model has  superior performance with less than half the parameters of Waveformer (0.498M vs 1.207M), demonstrating the efficiency of our architecture design.

\begin{table}[t!]
  \caption{Ablation study on FiLM layer placement. \textmd{Evaluation with 5 output channels, 1--5 targets in mixture.}}
  \label{tab:ablation_film}
  \vskip -0.1in
  \centering
  \setlength{\tabcolsep}{6pt}
  {\footnotesize
  \begin{tabular}{ l l c c }
    \toprule
    \textbf{Model} & \textbf{FiLM Placement} & \textbf{SNRi (dB)} & \textbf{SI-SNRi (dB)} \\
    \midrule
    Orange Pi & First block only & 11.76 $\pm$ 4.27 & 9.51 $\pm$ 5.43 \\
    Orange Pi & All blocks & \textbf{12.26 $\pm$ 4.38} & \textbf{10.16 $\pm$ 5.72} \\
    Orange Pi & All except first & 11.88 $\pm$ 4.27 & 9.77 $\pm$ 5.55 \\
    \midrule
    NeuralAids & First block only & 9.46 $\pm$ 3.81 & 5.73 $\pm$ 6.21 \\
    NeuralAids &  All blocks & \textbf{10.50 $\pm$ 4.15} & \textbf{7.27 $\pm$ 6.28} \\
    NeuralAids & All except first & 10.19 $\pm$ 4.01 & 7.09 $\pm$ 5.38 \\
    \bottomrule
  \end{tabular}
  }
  \vskip -0.2in
\end{table}

\subsubsection{Aurchestra's Model Evaluations} \label{sec:modelevals}
The above evaluation focused on a single target sound for fair comparison with prior work~\cite{semantichearing}. However, the key contribution of our architecture is its support for multiple target sources in the auditory environment. In this section, we evaluate how performance varies with the number of target sources in the mixture and the number of output channels in the model. Table~\ref{tab:orangepi_eval} presents a comprehensive comparison for different variants of our model running on the Orange Pi platform. Note that models with multiple output channels (e.g., 5 or 20) extract all specified targets simultaneously, whereas a single-output model requires separate inference for each target; in the latter case, we report the averaged metrics.

Several key observations emerge from this result. 
\squishlist
\item  Aurchestra maintains stable performance when increasing output channels from 1 to 5, with SNRi remaining around 9.5–12.0~dB for single-source mixtures. Performance drops sharply at 20 output channels, where SNRi falls from 11.99~dB (1 output) to 3.51~dB (20 outputs). 

\item Examining mixture complexity, Aurchestra maintains stable performance with multiple target sources in mixture, but declines with 4 or more. For 1-output, SI-SNRi drops from 11.27~dB (1 source) to 11.10~dB (3 sources), then more sharply to 9.79~dB (4 sources) and 9.33~dB (5 sources). This shows that while moderately complex mixtures are handled well, performance decreases when simultaneous targets exceed 5.

\item The 5-output configuration offers a good balance between flexibility and performance, achieving 9.56~dB SNRi with single-source mixtures while allowing users to select from 5 different sound classes simultaneously.

\squishend


\subsubsection{Accuracy–latency trade-off.} To make the accuracy-latency trade-off explicit, we vary the STFT window
configuration, i.e., chunk size ($L_C$), lookback ($L_B$), and lookahead
($L_F$), while keeping the total frame size fixed at 256 samples
($N_{\text{fft}} = L_C + L_B + L_F$). We evaluate two model sizes:
M ($D{=}32$, $H{=}64$, 0.50M parameters, our Orange Pi model) and
L ($D{=}64$, $H{=}128$, ${\sim}$1.5M parameters), both with FiLM applied
to all blocks and 5 output channels.

\begin{table}[t!]
  \caption{Accuracy-latency trade-off across model sizes and window configurations. \textmd{Evaluation with 5 output channels, 1-5 targets in mixture, FiLM on all blocks.}}
  \label{tab:latency}
  \vskip -0.1in
  \centering
  \setlength{\tabcolsep}{6pt}
  {\footnotesize
  \begin{tabular}{ l c c c }
    \toprule
    \textbf{Window ($L_C$/$L_B$/$L_F$)} & \textbf{Latency} & \textbf{M (0.50M)} & \textbf{L (${\sim}$1.5M)} \\
    \midrule
    6/6/4 (default) & 10\,ms & \textbf{10.16} & 9.17 \\
    4/8/4            & 8\,ms  & 8.79            & \textbf{10.58} \\
    8/4/4            & 12\,ms & 8.44            & 10.30 \\
    4/12/0 (causal)  & 4\,ms  & 8.14            & 9.49 \\
    \bottomrule
  \end{tabular}
  }
  \vskip -0.2in
\end{table}

Table~\ref{tab:latency} presents the results. The L model with a 4/8/4
window ($L_C{=}4$\,ms, $L_B{=}8$\,ms, $L_F{=}4$\,ms) achieves the best
SI-SDRi of 10.58\,dB at 8\,ms algorithmic latency, surpassing the
baseline M model (10.16\,dB at 10\,ms) in both accuracy (+0.42\,dB) and
latency ($-$2\,ms). This reveals a non-monotonic relationship between
latency and accuracy: lower latency can yield higher accuracy when the
latency budget is allocated appropriately.

Across configurations, allocating more samples to lookback (past context)
consistently outperforms larger chunk sizes. For the L model,
4/8/4 (10.58\,dB) outperforms both 8/4/4 (10.30\,dB, 12\,ms) and
6/6/4 (9.17\,dB, 10\,ms), despite having the lowest latency among the
three. This suggests that past temporal context provides more useful
information than larger analysis frames for real-time source separation.
Removing lookahead entirely (4/12/0, 4\,ms latency) incurs a meaningful
accuracy cost of $-$0.67 to $-$2.02\,dB, indicating that even minimal
future context is valuable. The M model does not benefit from window
optimization to the same extent (M~4/8/4: 8.79\,dB $<$ M~6/6/4:
10.16\,dB), suggesting that its performance is bounded by model capacity
rather than input information.


\subsubsection{Additional benchmarks (FUSS)} \label{secmain:fuss}
We also  evaluate on
the FUSS (Free Universal Sound Separation) eval set~\cite{wisdom2021s}, an
independently curated benchmark consisting of 1{,}000 mixtures with 1{,}445
foreground sources derived from FSD50K. Using a training-aligned class mapping that matches
our pipeline's AudioSet leaf-node ontology traversal, 180 of 1{,}445
foreground sources (12.5\%) across 15 of our 20 target classes are
 evaluable. Our model achieves SI-SDRi of $+$9.93\,dB on FUSS, within
0.22\,dB of our own evaluation set ($+$10.15\,dB). This close agreement
across independently constructed mixtures confirms that our extraction
model generalizes to unseen acoustic scenes without additional training.
Five classes (alarm clock, baby cry, cock-a-doodle-doo, music, ocean) have
zero matchable sources in FUSS due to taxonomy mismatches between FSD50K's
labeling granularity and our class definitions; detailed per-class analysis
is provided in~\xref{appendix:fuss}.


\vskip 0.03in\noindent{\bf Ablation study: FiLM layer placement.} As described in \xref{sec:arch}, FiLM layers inject conditioning information from the multi-hot encoding into the network. We compare three placement strategies: (1) applying FiLM only to the first block, (2) applying FiLM to all blocks, and (3) applying FiLM to all blocks except the first. Table~\ref{tab:ablation_film} presents results for both Orange Pi and NeuralAids models with 5 output channels.

For both models, applying FiLM to all blocks achieves the best performance. The Orange Pi model achieves 12.26~dB SNRi with FiLM on all blocks compared to 11.76~dB with FiLM only on the first block. The improvement is even more pronounced for the NeuralAids model, where applying FiLM to all blocks improves SI-SNRi from 5.73~dB to 7.27~dB.

\begin{figure}[t!]
\centering
\includegraphics[width=0.99\linewidth]{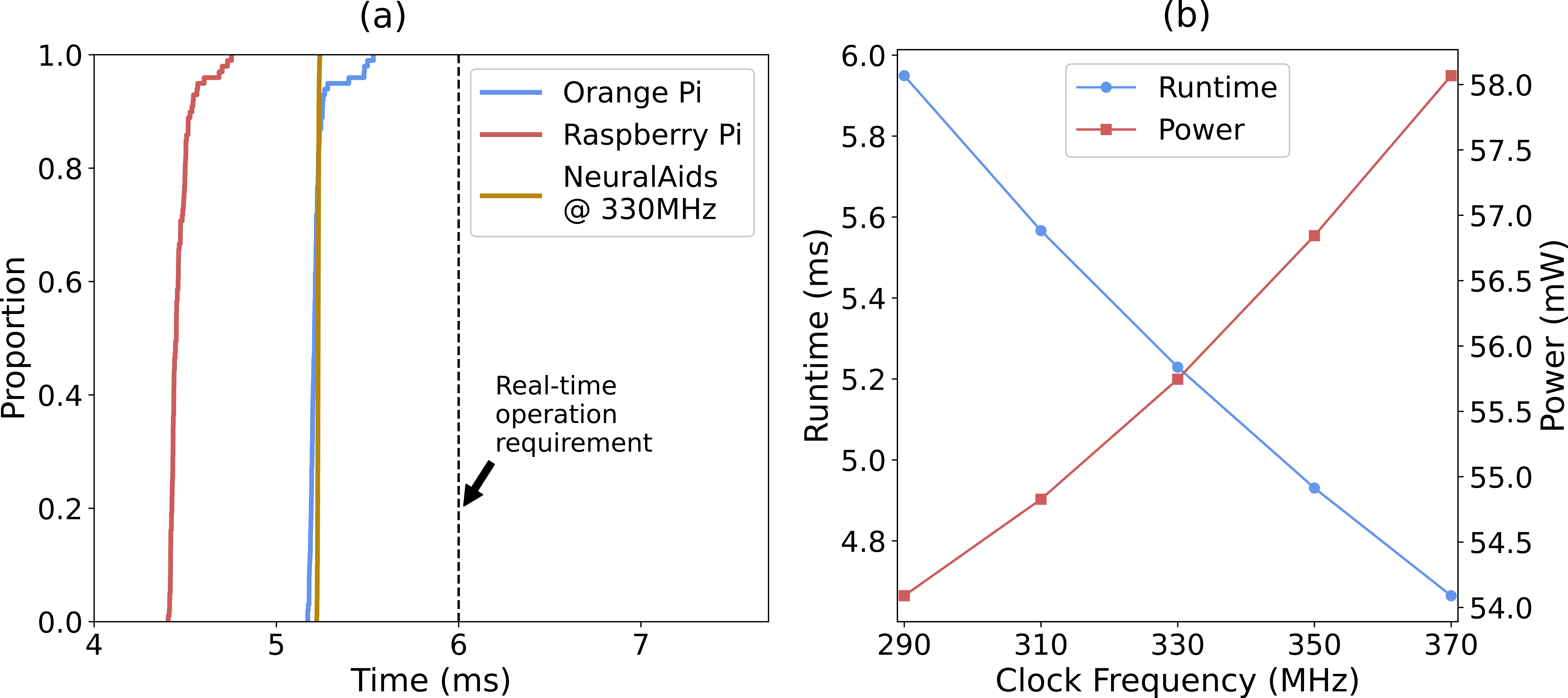}
\vskip -0.1in
\caption{\textmd{(a) CDF plots of the runtime of the three proposed networks on their respective deployment platforms. (b) Runtime and power consumption of running the NeuralAids model at different GAP9 clock frequencies.}}
\label{fig:hardware_results}
\vskip -0.2in
\end{figure}

\begin{figure*}[t!]
\centering
\includegraphics[width=0.8\linewidth]{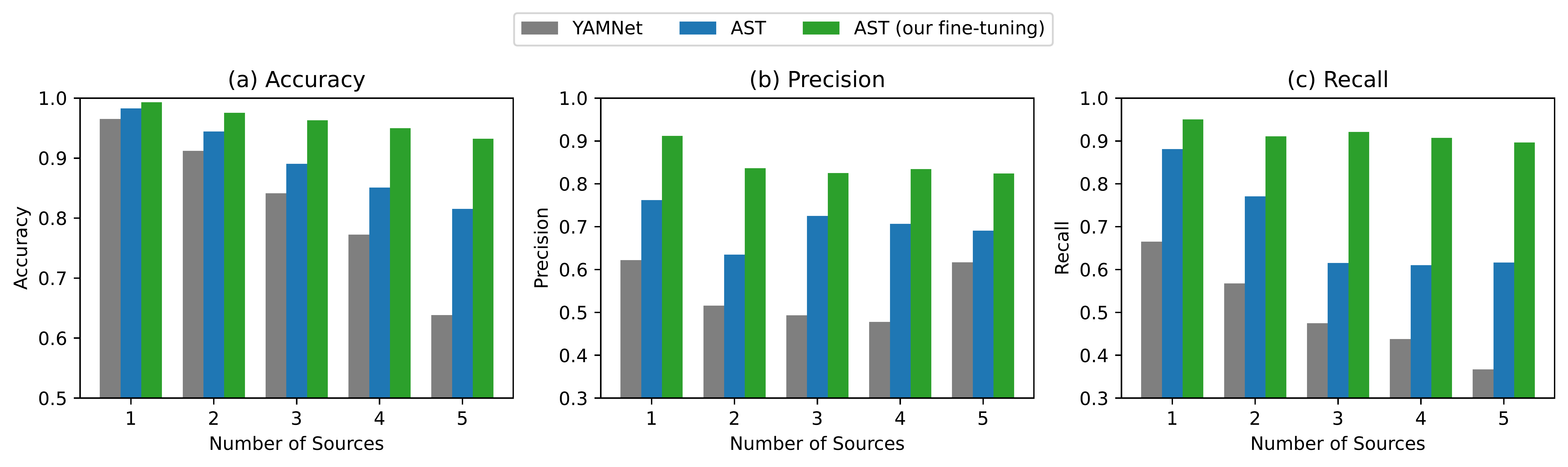}
\vskip -0.15in
\caption{Sound Event Detection performance comparison on 5-second audio segments. \textmd{We compare YAMNet, AST, and our fine-tuned AST model across different numbers of simultaneous sound sources. Our fine-tuned model maintains high performance even with 5 concurrent sources.}}
\label{fig:sed_comparison}
\vskip -0.2in
\end{figure*}

\subsubsection{Hardware evaluation}
We evaluate  our three network variants on their respective  hardware platforms. We use a chunk size of 6~ms, which implies the network runtime on each platform must not exceed 6~ms for real-time operation.

For the Orange Pi model, we use dynamic quantization for only the LSTMs. For the NeuralAids model, we quantize all weights and activations of all layers to \textsc{int8}, except for: 1) the encoder, 2) the decoder, 3) the LSTM,  addition, multiplication, activation functions, and recurrent states, and 4) all FiLM operations, which run in \textsc{bfloat16}.

We measure inference time on each of the three platforms  100 times. We discard the very first measurement which is typically much slower due to uninitialized cache.  Fig.~\ref{fig:hardware_results}a shows the CDF  of the inference time, which is well within the real-time constraints:  the mean inference time is 5.22~ms on the Orange Pi 5B, 4.47~ms on the Raspberry Pi 4B, and 5.23~ms on the GAP9 running at 330~MHz.

We also measure the inference time and power consumption of our  NeuralAids model as a function of the device's clock rate. We observe in Fig.~\ref{fig:hardware_results}b a slight power reduction as we decrease the clock rate, at the cost of higher inference time. At 290~MHz, the neural network runs in real-time on NeuralAids and consumes just 54.1~mW.



\subsection{Benchmarking Sound Event Detection}

\subsubsection{Evaluation Metrics}

Our evaluation of Aurchestra's Sound Event Detection (SED) model consists of two main components: classification performance and system performance. The former assesses how accurately the model detects and classifies sounds in the current environment, while the latter measures inference time for real-time operation.
\squishlist
\item {\it Accuracy.} This measures the proportion of correct predictions and is a key metric. However, in the multi-label classification problem addressed in this work, class imbalance may exist, making accuracy alone insufficient to fully reflect the model's actual performance. Thus we conduct a comprehensive evaluation using additional metrics.

\item{\it Precision.} This measures the proportion of  correct positive instances among those predicted as positive for a specific sound class. High precision indicates fewer false positives, which minimizes the display of non-existent sound labels. 

\item{\it Recall.}
This measures the proportion of correct positive instances that the model correctly detected. High recall indicates fewer false negatives, which ensures that all available sound labels are displayed on the interface without omission.

\item{\it F1-score.} This is the harmonic mean of precision and recall, capturing their balance in a single value. We selected the classification threshold that maximized F1 on the validation set and applied it to evaluate test performance.

\item{\it Runtime.} In our system, the SED model runs on the smartphone. Runtime measures the time to process each  audio segment to assess real-time capability.

\squishend

\subsubsection{Model Comparison}\label{sec:SED:comp}

We benchmark the SED models using a test dataset of audio  consisting of one or more of our 20  target sound classes.  We compare three models:

\squishlist
\item {\it YAMNet:} A  convolutional neural network pre-trained on AudioSet, designed for efficient audio event classification. YAMNet uses MobileNet-v1 architecture  and operates on mel-spectrogram inputs.
\item {\it AST:} Audio Spectrogram Transformer pre-trained on AudioSet. AST applies a Vision Transformer (ViT) architecture to audio spectrograms, achieving strong performance on audio classification tasks.
\item {\it AST (Our fine-tuning):} The AST model fine-tuned  following the procedure described in~\xref{sec:finetuning}, with differential learning rates for the encoder and classifier.
\squishend

We evaluate each model across two dimensions: (1) the number of simultaneous sound sources in the mixture (1 to 5 sources), and (2) the duration of the audio segment (2s to 10s). This  evaluation reveals how model performance degrades as the acoustic scene becomes more complex.

Fig.~\ref{fig:sed_comparison} shows the classification performance of the three models on 5-second audio segments with varying numbers of sound sources.   Our fine-tuned AST model consistently outperforms both the baseline AST and YAMNet across all three metrics and mixture conditions, highlighting the importance of our finetuning procedure. For single-source detection, all three models achieve high accuracy (above 96\%). However, as the number of sources increases, performance differences become more pronounced. With five simultaneous sources, YAMNet's accuracy drops to 63.8\%, while the baseline AST was 81.5\% while our fine-tuned model achieves 93.2\%.

The improvement is particularly notable in precision and recall metrics. Our fine-tuned model maintains precision above 82\% even with 5 sources, compared to 69.1\% for baseline AST and 61.7\% for YAMNet. Similarly, recall remains above 89\% for our model versus 61.6\% for AST and 36.7\% for YAMNet with 5 sources. These results demonstrate that fine-tuning on our dataset significantly improves the model's ability to detect multiple target sounds without increasing false positives or missing actual sound events.

The detailed performance across all duration and source combinations is presented in~\xref{appendix:sed}. As expected, performance generally improves with longer audio segments, as more temporal context allows for better sound event detection. With 10-second segments and a single source, the model achieves 99.4\% accuracy, 91.9\% precision, 95.1\% recall, and 93.5\% F1-score.

The performance degradation with shorter segments is more pronounced when multiple sources are present. For instance, with a single source, reducing duration from 10s to 2s decreases F1-score from 93.5\% to 88.6\%. However, with 5 sources, the same duration reduction causes F1-score to drop from 87.4\% to 74.7\%. This suggests that shorter observation windows make it harder to disambiguate overlapping sounds. Notably, even in the most challenging condition (2-second segments with 5 sources), our fine-tuned model achieves 88.1\% accuracy and 74.7\% F1-score. 

To complement the aggregate metrics in Fig.~\ref{fig:sed_comparison}, we
analyze per-class detection patterns. The inter-class confusion matrix
for single-source scenes (Fig.~\ref{fig:confusion_matrix} in
Appendix~\ref{appendix:sed}) shows that the SED module achieves recall
above 0.9 for most classes, with off-diagonal confusion concentrated
among impulsive percussive sounds: hammer, door knock, gunshot, and glass
breaking share short broadband energy profiles that make them mutually
confusable even in isolation.

As the number of simultaneous sources increases,
Table~\ref{tab:perclass_f1} (Appendix~\ref{appendix:sed}) shows that
macro F1 drops from 0.923 (\#sources$=$1) to 0.847 (\#sources$=$5), an
8 percentage-point decline, with 15 of 20 classes maintaining F1 above
0.80. The sharpest drop occurs at the transition from single- to
multi-source detection (\#sources$=$1$\to$2, $\Delta$F1$=-$0.051), with
near-plateau behavior thereafter. Classes with acoustically distinctive
signatures such as alarm clock (periodic tone), toilet flush (broadband
rush), and music (harmonic structure) remain stable across all
conditions, while impulsive percussive sounds (glass breaking, gunshot)
show the largest degradation.

\subsubsection{SED runtime evaluation}\label{sec:sed:runtime}
We evaluate the inference runtime of our fine-tuned AST model across smartphones to assess its suitability for real-time operation. Since the SED model runs on the smartphone and periodically updates the detected sound labels on the application interface, the inference time determines how frequently the system can refresh the available sound classes for user selection. 

\vskip 0.03in\noindent{\bf Evaluation procedure.}
We measure the end-to-end inference time for processing 5-second audio segments on three iPhone models spanning different hardware generations: iPhone 17 Pro (latest generation), iPhone 15, and iPhone 12 Pro. Runtime is measured by recording timestamps before and after model execution, and we report CDFs  over 100 inference iterations to capture runtime variability.

\begin{figure}[t!]
\centering
\includegraphics[width=0.8\linewidth]{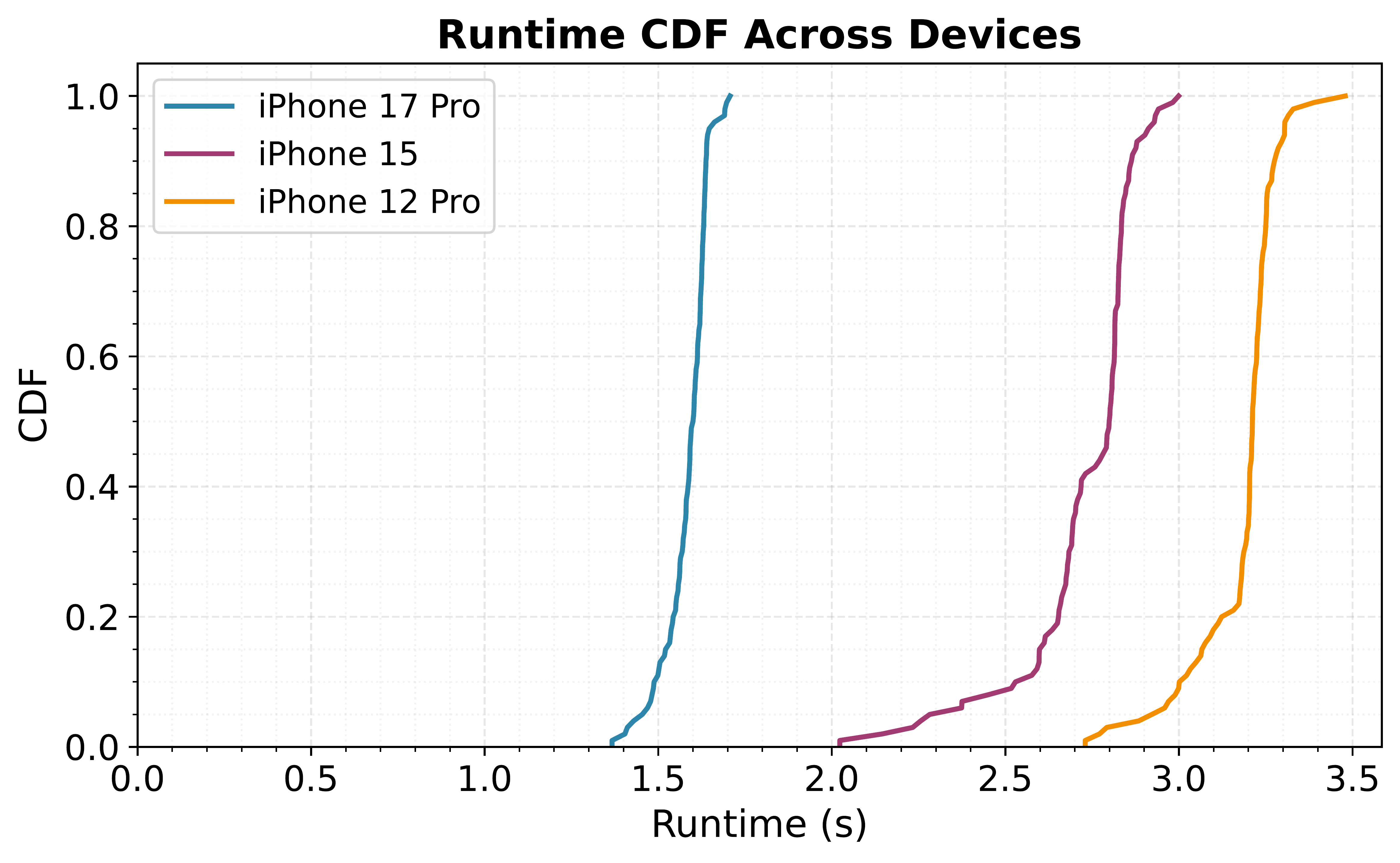}
\vskip -0.1in
\caption{Runtime CDF of the fine-tuned AST model across different iPhone platforms. \textmd{All platforms complete inference faster than the 5-second audio segment duration.  }}
\label{fig:sed_runtime}
\vskip -0.2in
\end{figure}

\vskip 0.03in\noindent{\bf Results.} Fig.~\ref{fig:sed_runtime} shows the runtime CDF for the three platforms. The iPhone 17 Pro is fastest, with a median runtime of around 1.6s and 95th percentile under 1.8s. The iPhone 15 and iPhone 12 Pro have median runtimes of  2.9s and 3.3s, respectively. Since all devices process 5s audio segments within 5s, the SED model  runs in real time on all tested devices.

\begin{figure*}[t!]
\centering
\includegraphics[width=0.9\linewidth]{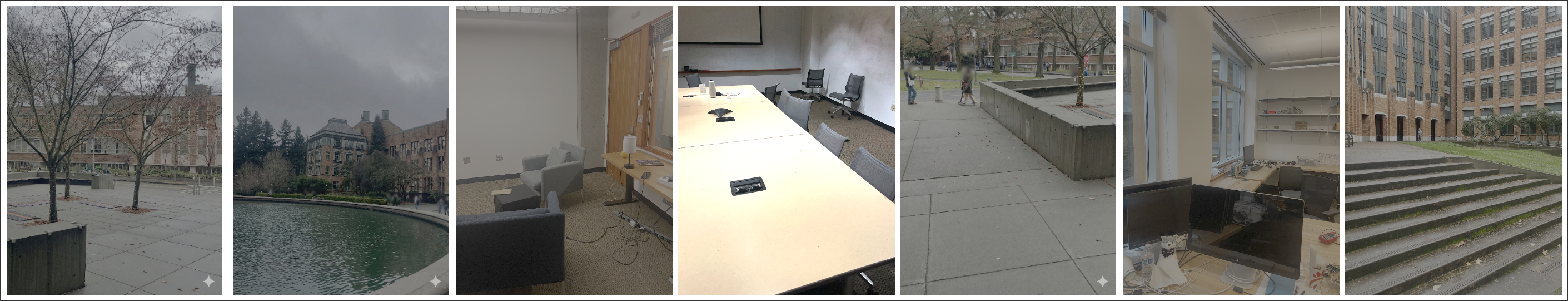}
\vskip -0.15in
\caption{In-the-wild  scenarios. \textmd{The wearer and sound sources were free to move, and head rotation was uncontrolled.}}
\label{fig:wild_environments}
\vskip -0.18in

\end{figure*}

\subsection{In-The-Wild Evaluation}

To evaluate our system's real-world performance, 5 individuals (3 female and 2 male) wore
our headsets and collected audio recordings in  diverse environments. As shown in Fig.~\ref{fig:wild_environments}, the recording locations included indoor settings like offices with ambient chatter and typical workplace noises (keyboard typing, conversations), as well as outdoor locations including busy streets with traffic noise and natural environments like parks with multiple sound sources. In all recording scenarios, the position and movement of sound sources were uncontrolled and reflective of real-world conditions. 

Since some sound classes were more common in natural environments than others, our in-the-wild evaluation focused on a subset of target classes that most frequently appeared in our recordings. Each recording contained 1-2 target sounds with background urban or indoor ambient noise persisting throughout the recording duration. 

\vskip 0.05in\noindent{\bf Evaluation protocol.}
As clean, sample-aligned clean  signals are unavailable for real-world recordings, the above metrics cannot be used. So, we conducted a listening study to obtain Mean Opinion Scores (MOS) for sound extraction quality. 17 participants (11 male, 6 female) rated each sample on three 5-point scales: (1) target-sound clarity (1 = extremely distorted, 5 = not distorted), (2) background-noise intrusiveness (1 = extremely intrusive, 5 = not noticeable), and (3) overall listening experience (1 = bad, 5 = excellent).

\begin{figure}[t!]
\centering
\includegraphics[width=\linewidth]{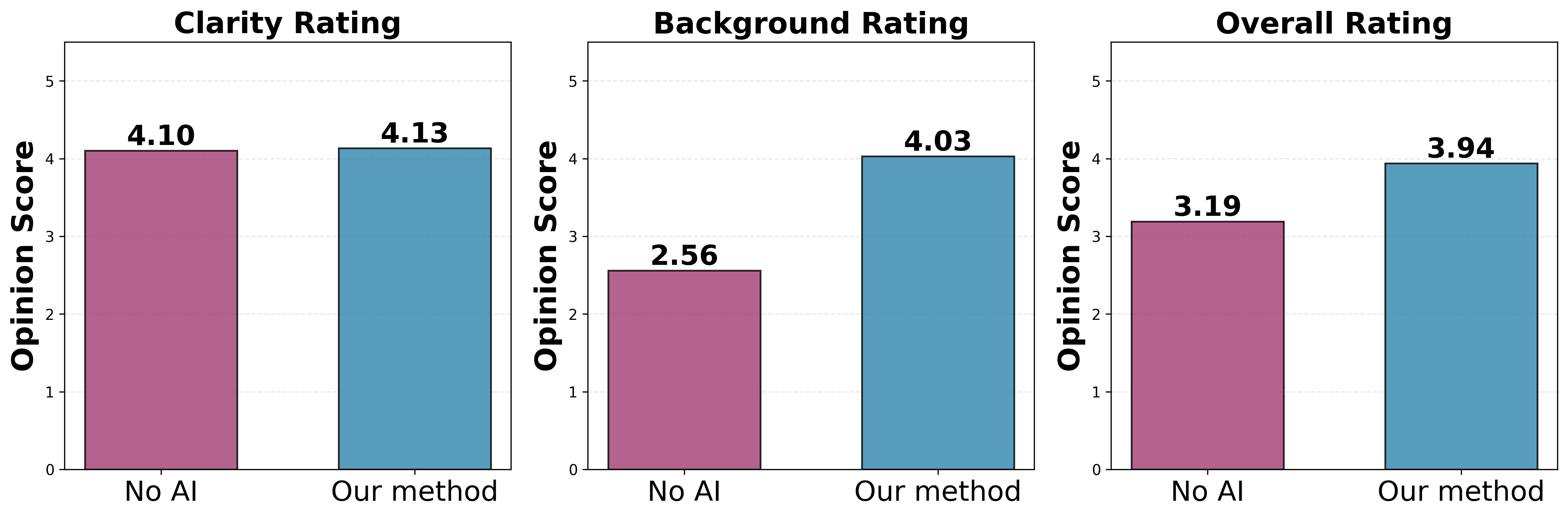}
\vskip -0.1in
\caption{User listening study results comparing Aurchestra against the No AI baseline. \textmd{ Aurchestra achieves substantial improvements in background noise suppression (+1.54) and overall experience (+0.95) while maintaining target sound clarity. N=17 participants.}}
\label{fig:rating_comparison}
\vskip -0.15in
\end{figure}

\vskip 0.05in\noindent{\bf Results.}
Fig.~\ref{fig:rating_comparison} shows ratings comparing Aurchestra with the No-AI baseline. Our system yields substantial improvements in background-noise suppression  and overall listening experience, while maintaining comparable target-sound clarity, indicating that the extraction process preserves perceptual quality without introducing noticeable distortion.




Fig.~\ref{fig:per_class_clarity} shows clarity ratings by target sound class, revealing variation across sound types, with all classes scoring above 3.4 on the 5-point scale. Impulsive, distinctive sounds achieved the highest clarity: Alarm Clock (4.59), Hammer (4.47), and Toilet Flush (4.29). Their sharp transients and distinct spectral signatures make them easier for the extraction network to isolate without audible artifacts.


Sounds with broader spectral content or complex temporal patterns received lower clarity ratings: Birds Chirping (3.50), Computer Typing (3.82), and Car Horn (3.91). Birds Chirping is challenging due to its wide frequency range and overlap with background components, while Computer Typing’s rapid clicks can coincide with other percussive noises, hindering clean separation. Speech achieved a moderate score of 4.00, indicating acceptable quality but showing room for improvement, especially in multi-speaker scenarios.



\subsection{Dynamics Interface Evaluation}

We evaluate the interface design via task completion time  and System Usability Scale (SUS) questionnaire responses.

\subsubsection{Interface comparison}

We compared two  interfaces:

\squishlist
\item {\it Static Interface:} Displays an alphabetical list of all 20 predefined sound categories. Users must scroll through the entire list to find and select the sounds they are currently hearing.
\item {\it Dynamic Interface:}  Automatically detect sounds and display only the identified  categories  on the app interface. Users select target sounds from this filtered, context-aware list. 
\squishend


Seven participants used both interfaces in a within-subjects design. A total of 10 mixtures containing 3 target sounds were looped across both interface.  The interfaces were presented in a random order, with no repetition of mixtures across the two interfaces.  For each interface, participants selected two sounds present in their environment, and we measured the time from interface presentation to successful selection.

\begin{figure}[t!]
\centering
\includegraphics[width=0.9\linewidth]{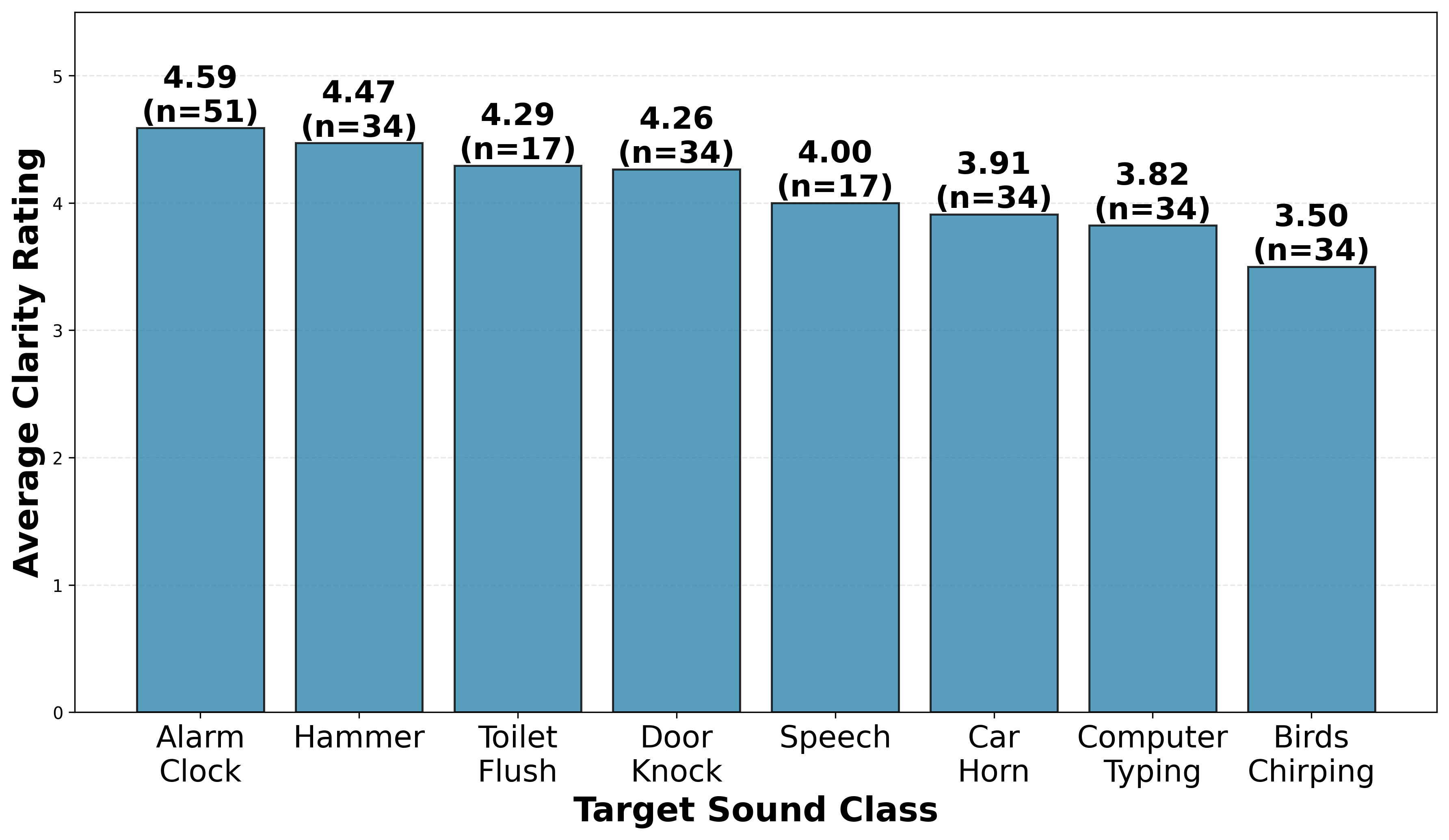}
\vskip -0.15in
\caption{Aurchestra's per-class clarity ratings. }
\label{fig:per_class_clarity}
\vskip -0.2in
\end{figure}

\begin{figure}[t!]
\centering

\includegraphics[width=0.45\linewidth]{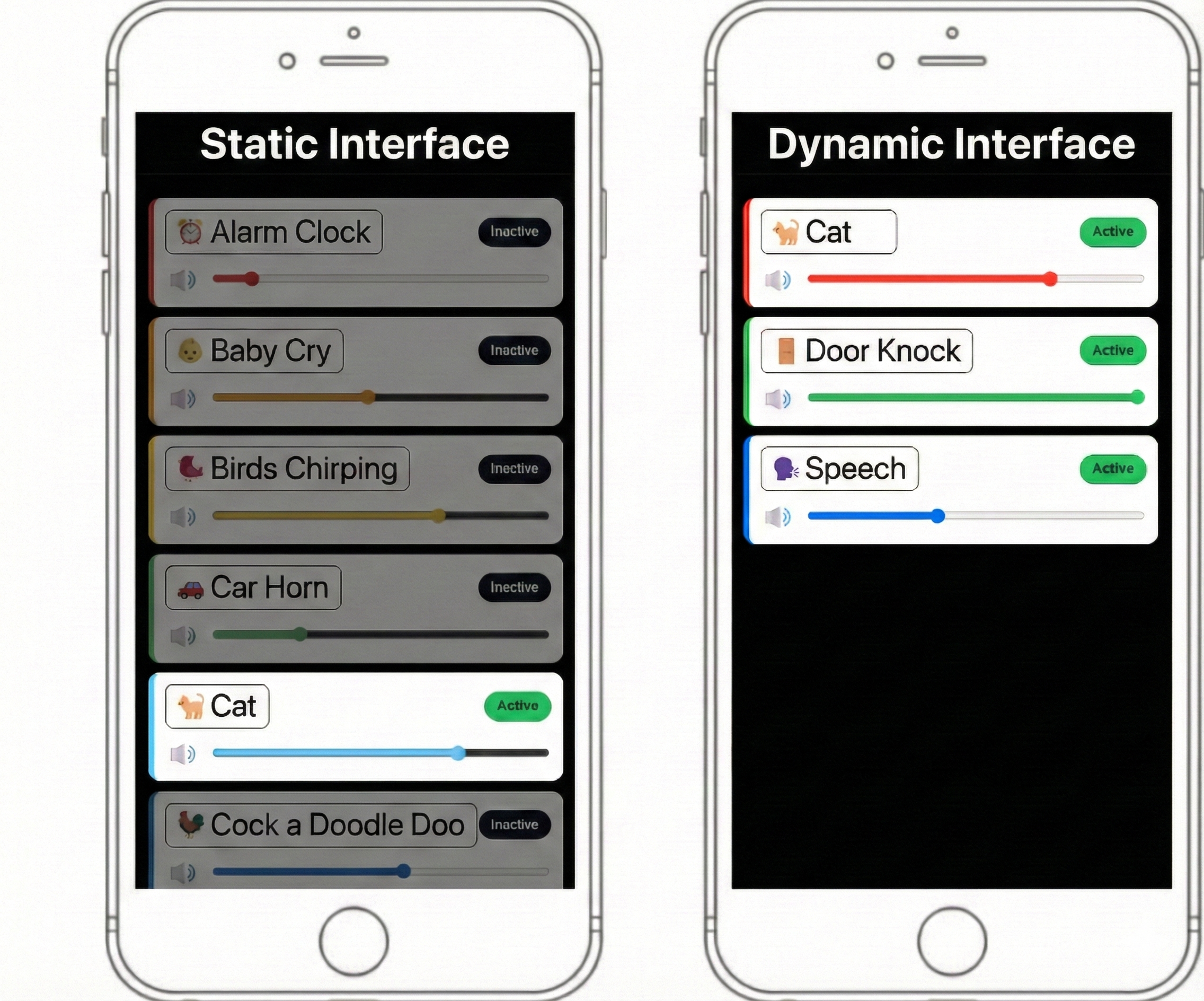}
\includegraphics[width=0.45\linewidth]{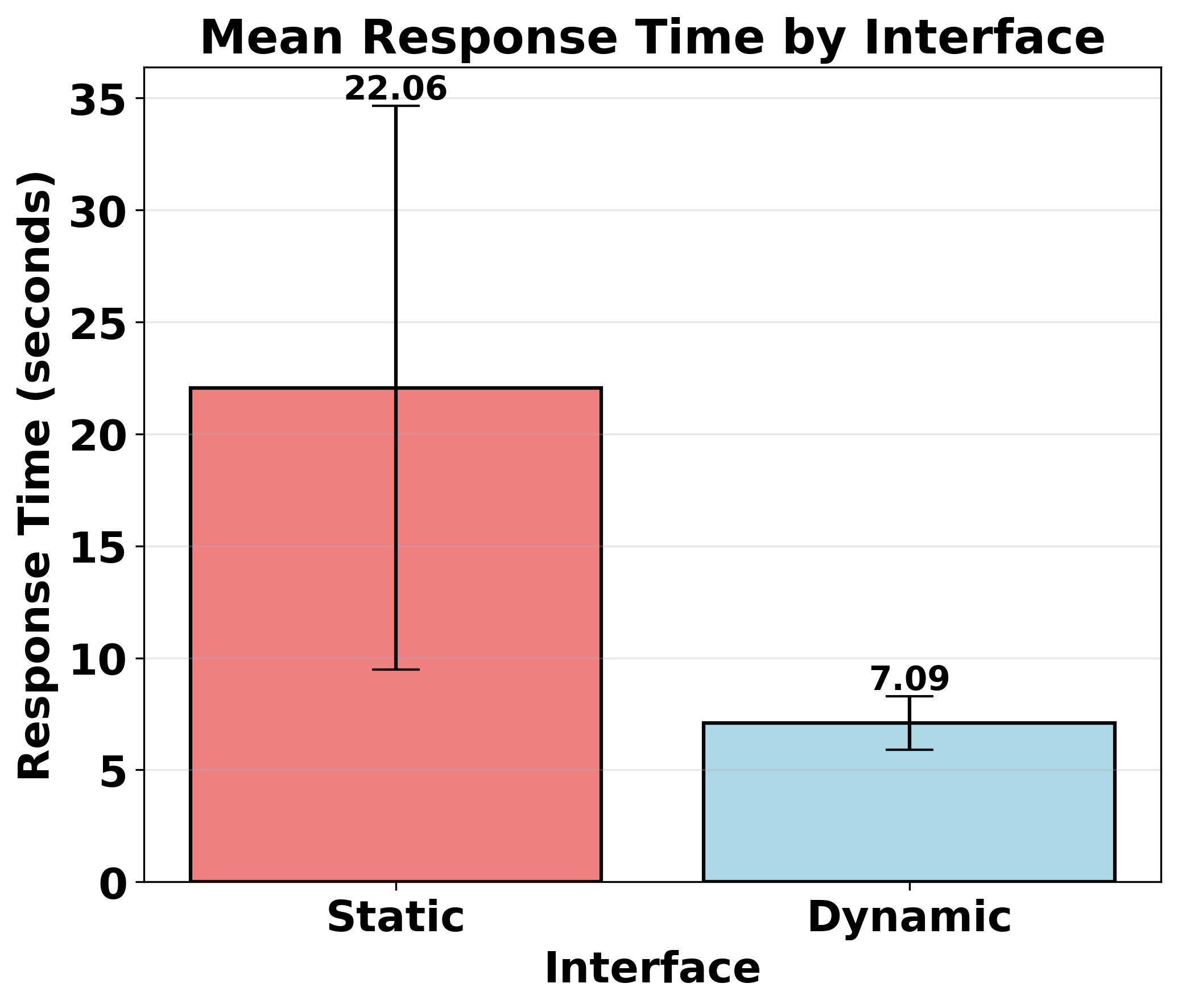}
\vskip -0.1in
\caption{Response times for  sound selection. \textmd{Our dynamic interface reduces selection time by 67.9\% compared to the static interface by displaying only detected sounds rather than all 20 categories. Error bars show standard deviation.  }}
\label{fig:response_time}
\vskip -0.15in
\end{figure}

Fig.~\ref{fig:response_time} shows that the dynamic interface achieved a significant reduction in the  mean selection time demonstrating that showing only relevant sound options greatly accelerates interaction. The static interface forces users to scroll through all 20 sound classes, increasing effort and search time, whereas the dynamic interface displays only detected sounds, aligning options with users’ needs and enabling faster, more intuitive selection. Variance in selection time was also lower for the dynamic interface. The static interface exhibited high variability (SD=9–35s), likely reflecting differences in users’ familiarity with sound labels and their search strategies in a long list.



\begin{figure}[t!]
\centering
\includegraphics[width=\linewidth]{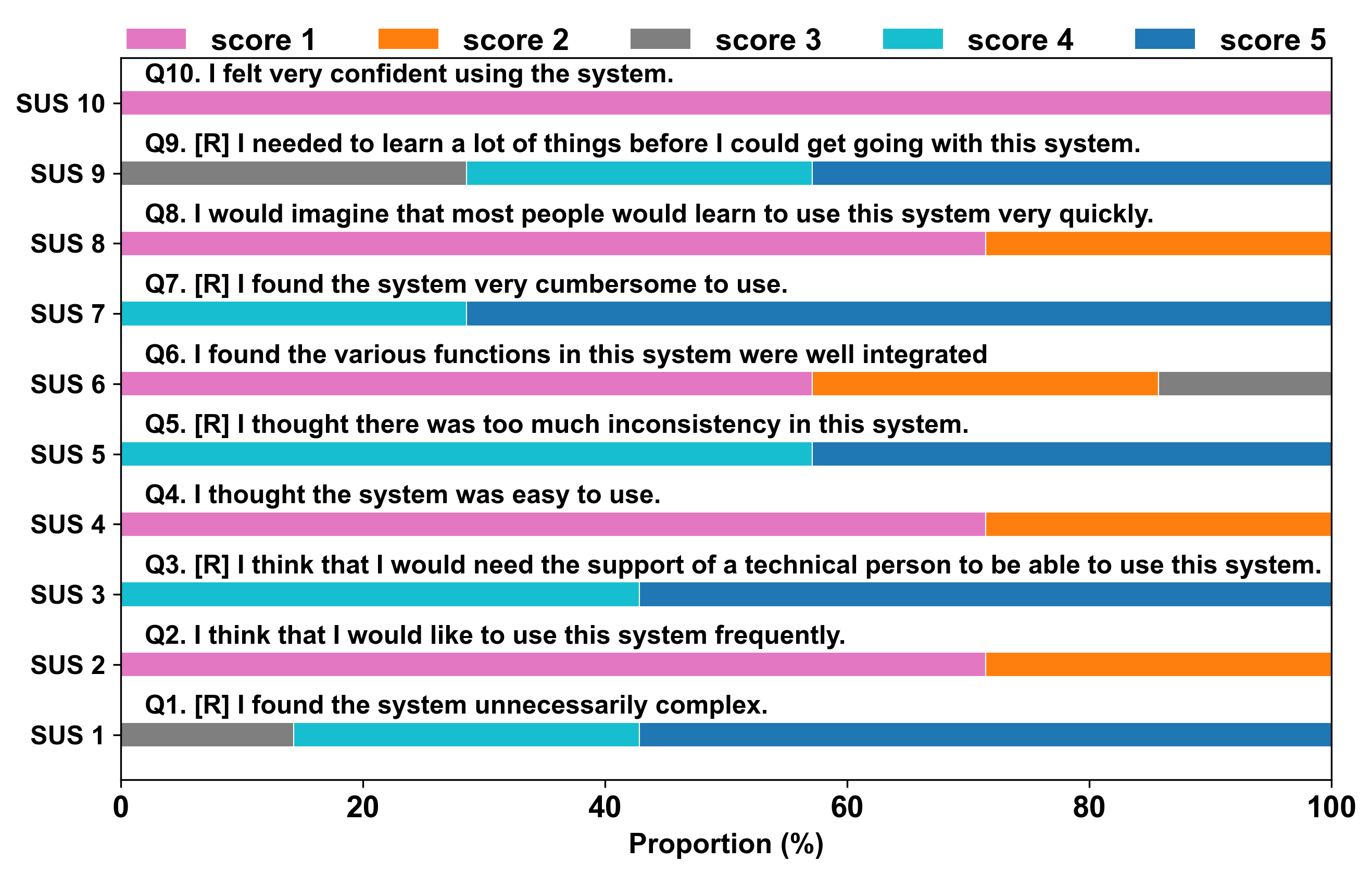}
\vskip -0.1in
\caption{System Usability Scale (SUS) evaluation. \textmd{Distribution of participant responses for each SUS item on a 5-point scale. Items marked with [R] are reverse-scored.  }}
\label{fig:sus_results}
\vskip -0.15in
\end{figure}

We also administered the standard 10-item SUS questionnaire to evaluate the overall usability of our system. The  participants rated each statement on a 5-point Likert scale (1 = Strongly Agree, 5 = Strongly Disagree for positively worded items; reversed for negatively worded items marked with [R]). Fig.~\ref{fig:sus_results} presents the distribution of responses for each SUS item. Results indicate strong positive usability ratings across learnability, ease of use, and user confidence.

\section{Limitations and Discussion}

{\it Dynamic interface improvements.} Our current  prototype  focuses on the technical  feasibility of identifying and displaying sounds detected in the immediate environment, which raises an important question about how users can select sounds they heard previously or expect to encounter soon. Users may want to configure their preferences in advance, e.g., choosing to focus on “speech” before entering a meeting room or selecting “alarm” or “car horn” as always-important sounds regardless of current conditions. One potential solution is to incorporate a “recently heard sounds'' list, allowing users to quickly re-select sounds detected recently. Another option is to introduce a “favorites” or “preferred sounds” feature, enabling users to pin specific classes that should always remain visible in the interface. We could also consider a predictive model that anticipates likely future sounds based on context, such as location, time of day, or user activity. Alternatively, a hybrid interface could combine dynamic detection with a collapsible panel listing all sound classes, giving users full control without overwhelming the default view.  Further user-centric research is needed to determine which combination of these approaches best aligns with usability goals, system complexity, and real-world user behavior.

\vskip 0.03in\noindent{\it Sound extraction improvements.} Our proof-of-concept prototype currently operates on a finite set of 20 sound classes. While this predefined taxonomy is adequate for demonstrating feasibility, scaling to a larger and more flexible class inventory or adopting hierarchical or open-set classification strategies would allow the device to recognize novel or rare sound categories beyond those seen during training. Realizing this on low-power hardware  will require mechanisms for downloading, adapting, or generating custom on-device models tailored to novel sounds without exceeding compute or memory constraints. Some classes are also inherently more difficult to separate because they share overlapping acoustic characteristics. For example, both music and speech contain vocal and harmonic components, and music can also resemble alarms or bird chirps in certain frequency bands. Developing audio embeddings that are tailored to these particularly challenging, acoustically similar classes may offer a path toward improved separation performance.

\vskip 0.03in\noindent{\it Inter-class confusion in detection.} This challenge extends beyond spectral overlap to temporal similarity.
Our inter-class confusion analysis
(Fig.~\ref{fig:confusion_matrix} in Appendix~\ref{appendix:sed})
reveals that impulsive, short-duration sounds, such as hammer strikes,
door knocks, gunshots, and glass breaking, are particularly susceptible
to mutual confusion when multiple sources are present, as they share
similar broadband temporal envelopes. In contrast, classes with
distinctive spectral signatures (e.g., alarm clock, music, toilet flush)
maintain high detection accuracy even in dense five-source scenes, with
F1 scores above 0.90. This pattern is consistent with the per-class
clarity ratings in Fig.~\ref{fig:per_class_clarity}, where impulsive
sounds achieve high individual clarity but are harder for the SED module
to distinguish from one another. These findings suggest that future work
on impulsive sound disambiguation, for example, through temporal
fine-structure analysis or onset-pattern features, could meaningfully
improve class-level detection in complex acoustic scenes.

\vskip 0.03in\noindent{\it Beyond per-class volume control.} Modern audio editing tools offer  many more effects, like equalization, reverb, or modulation, all of which can be applied on a per-class basis. For example, a useful effect for people with high-pitch hearing loss is to apply a downward pitch shift to sound classes with high-frequency components -- such as the ringing of an alarm clock -- effectively lowering the pitch to a range that is more audible to them. Incorporating such existing signal processing effects would give users more freedom to customize their soundscapes.

\vskip 0.03in\noindent{\it Open Problems.} Research on enhanced hearing has largely progressed along two parallel threads: systems that classify general environmental sounds and those that provide fine-grained control over speech sources. Semantic hearing \cite{semantichearing} and Aurchestra   fall into the first category, where speech is represented as a single broad class within a wider acoustic taxonomy. In contrast, prior work~\cite{lookoncetohear,soundbubble} focus on enabling users to selectively attend to specific talkers. These approaches  treat all non-speech sounds as background noise and do not reason about the broader acoustic scene. Bridging these two research threads, by creating a unified system that can jointly understand diverse  sounds while also providing speaker-level selectivity, remains an open  problem. 


\section{Conclusion}

We present Aurchestra, the first system to enable fine-grained, multi-class soundscape control for  resource-constrained hearables. Aurchestra allows users to independently mix sounds, transforming the soundscape from a single undifferentiated stream into something programmable.  By giving listeners the ability to sculpt their auditory world, Aurchestra takes an important step toward a new generation of intelligent hearables devices that  understand the user's  environments, and enable richer, more personalized listening experiences.


\balance
\bibliographystyle{ACM-Reference-Format}
\bibliography{refs}

\appendix

\begin{figure*}[t!]
\centering
\includegraphics[width=0.9\linewidth]{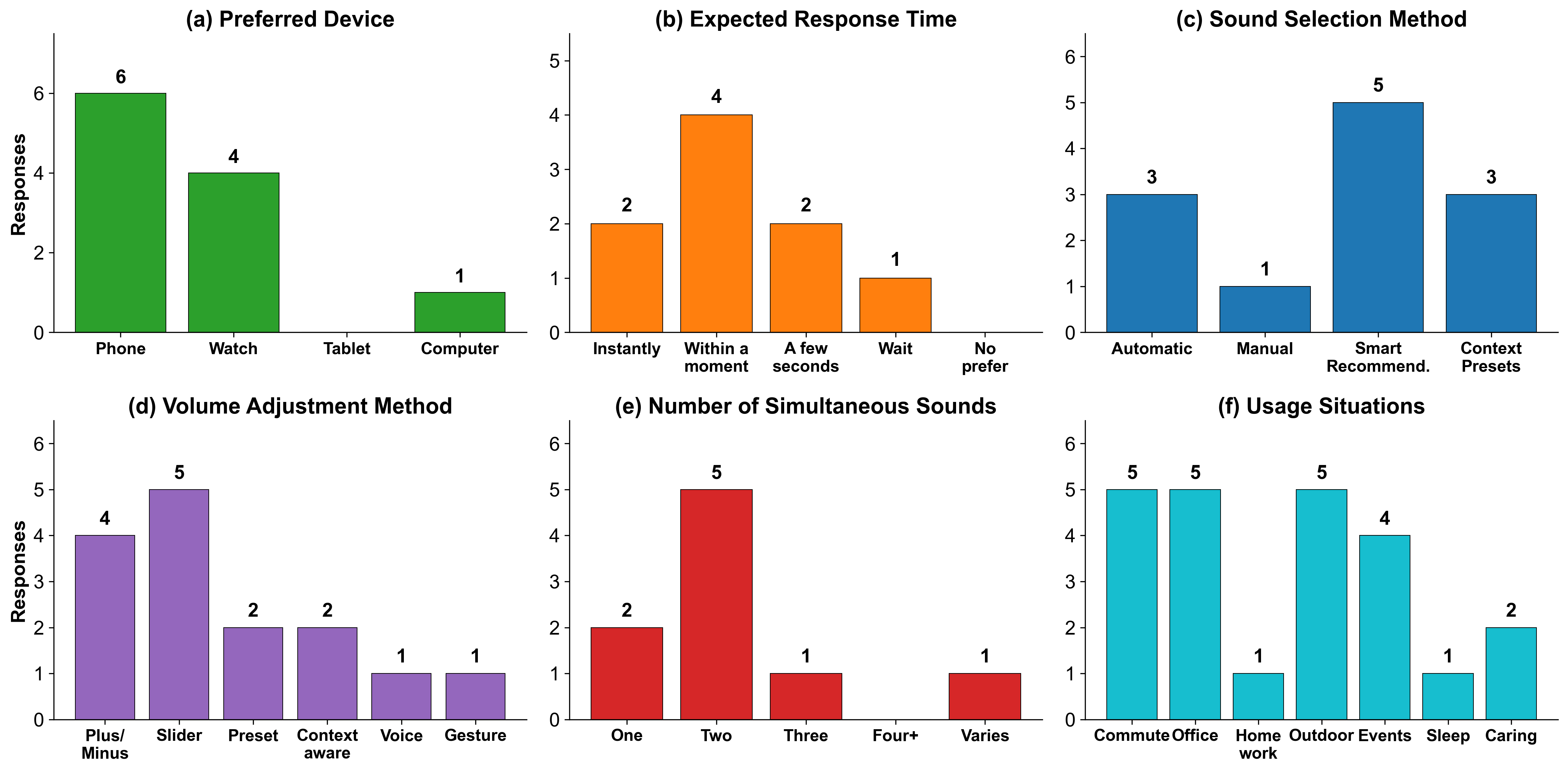}
\vskip -0.1in
\caption{User preferences survey results (N=7). \textmd{(a) Preferred device. (b) Expected response time. (c) Sound selection method. (d) Volume adjustment method. (e) Number of simultaneous sounds. (f) Usage situations.}}
\label{fig:survey_results}
\vskip -0.15in
\end{figure*}


\section{User Preferences Survey}

The  participants  in our user study also participated in a  survey to understand their preferences  for sound filtering applications. The participants were allowed to pick multiple options for each of these questions.  


Fig.~\ref{fig:survey_results} shows that the participants preferred mobile and wearable devices for using a sound filtering interface. Regarding response time expectations, the majority of participants (57.1\%) found a brief pause acceptable, while 28.6\% expected instant response with no noticeable delay. Only one participant prioritized accuracy over speed, suggesting that most users value responsive interaction.


For sound selection methods, Smart Recommendations, where the system suggests sounds and users approve or adjust, was the most preferred approach (71.4\%), followed by Automatic detection (42.9\%) and Context-based Presets (42.9\%). Only one participant preferred fully manual control. This preference distribution validates our dynamic interface design, which automatically detects and recommends sounds present in the environment rather than requiring users to manually search through all categories.


Most participants (71.4\%) preferred focusing on two sounds simultaneously (a primary and a secondary), while 28.6\% preferred focusing on only one. This reinforces the importance of supporting multi-target sound extraction rather than restricting the system to single-source enhancement. The most common situations where participants wanted to focus on specific sounds were commuting (71.4\%), working in an office (71.4\%), and exercising or outdoor activities (71.4\%), followed by attending public events (57.1\%).




Finally, open-ended responses revealed that speech and conversation were the most frequently desired target sounds. Participants also expressed interest in hearing music, nature sounds (e.g., birds, ocean), and safety-related cues such as alarms, traffic honks, and door knocks. Unwanted sounds overwhelmingly included traffic and urban noise (e.g., bus engines, construction noise, street noise). Indoor background sounds such as HVAC systems, vacuum cleaners, and typing noises were also commonly disliked. Human-generated distractions like babble, crying, chewing, and coughing appeared across multiple responses. These preferences closely align with Aurchestra’s predefined 20 target sound classes and the interfering classes used for suppression.


\section{Additional SED Evaluation}
\label{appendix:sed}

This appendix presents additional sound event detection results.

\subsection{Inter-class Confusion Matrix}

\begin{figure*}[t!]
\centering
\includegraphics[width=0.75\linewidth]{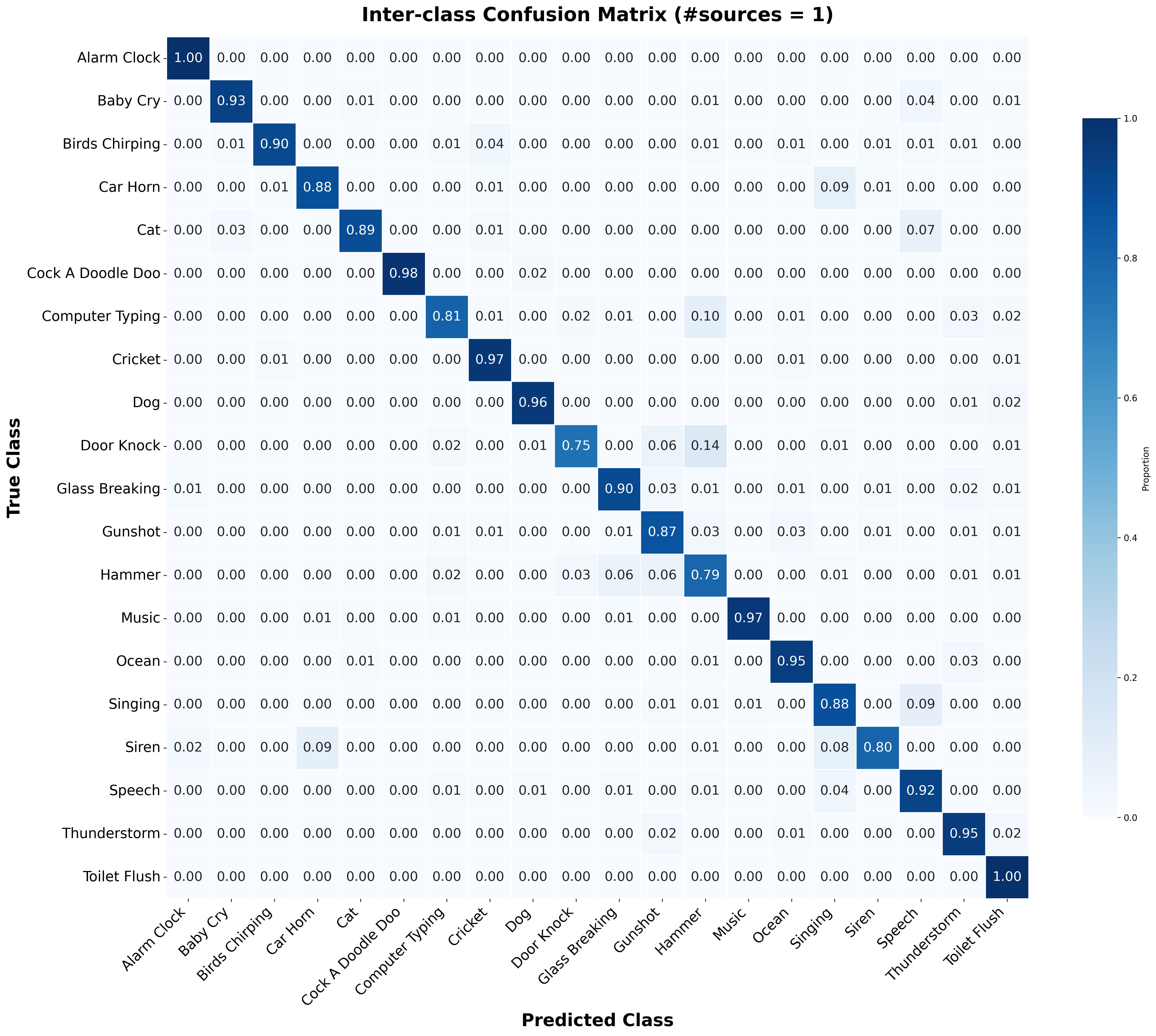}
\vskip -0.1in
\caption{Inter-class confusion matrix for single-source detection (\#sources$=$1). \textmd{Each cell shows the fraction of samples where the true class (row) is predicted as another class (column). Diagonal values indicate per-class recall. Confusion is concentrated among impulsive percussive sounds (hammer, door knock, gunshot, glass breaking), which share similar short broadband energy profiles.}}
\label{fig:confusion_matrix}
\vskip -0.1in
\end{figure*}

Fig.~\ref{fig:confusion_matrix} shows the inter-class confusion matrix for single-source scenes (\#sources$=$1). Each cell shows the fraction of samples where the true class (row) is predicted as another class (column); diagonal values indicate per-class recall. Most classes achieve recall above 0.9, with off-diagonal confusion concentrated among impulsive percussive sounds: hammer, door knock, gunshot, and glass breaking share short broadband energy profiles that make them mutually confusable even in isolation.

\subsection{Per-class F1 Scores}

Table.~\ref{tab:perclass_f1} presents per-class F1 scores as the number of simultaneous sound sources increases from 1 to 5. We define \#sources as the number of distinct target sound classes simultaneously present in the mixture. Overall macro F1 drops from 0.923 (\#sources$=$1) to 0.847 (\#sources$=$5), an 8 percentage-point decline, with 15 of 20 classes maintaining F1 above 0.80. The sharpest drop occurs at the transition from single- to multi-label detection (\#sources$=$1$\to$2, $\Delta$F1$=-$0.051), with near-plateau behavior thereafter (\#sources$=$3$\to$5, $\Delta$F1$<$0.015 per step).

The table is grouped into three tiers separated by horizontal rules: (1)~\emph{stable} classes (top 8, F1 drop $<$0.06) with acoustically distinctive signatures such as periodic tones (alarm clock) or unique spectral patterns (toilet flush); (2)~\emph{moderate} classes (middle 8, drop 0.06--0.12) consisting of broadband or environmental sounds with overlapping frequency bands; and (3)~\emph{vulnerable} classes (bottom 4, drop $>$0.12) consisting entirely of impulsive percussive sounds that share short broadband energy bursts.

\begin{table*}[t!]
  \caption{Per-class F1 scores across increasing numbers of simultaneous sound sources (\#sources). \textmd{Evaluated on 5-second segments with optimal per-class thresholds. \#sources denotes the number of distinct target sound classes simultaneously present in the mixture. Bold indicates F1 $>$ 0.90 at \#sources$=$5. Classes are grouped into three performance tiers (separated by horizontal rules).}}
  \label{tab:perclass_f1}
  \vskip -0.1in
  \centering
  \setlength{\tabcolsep}{5pt}
  {\footnotesize
  \begin{tabular}{ l c c c c c c }
    \toprule
    \textbf{Class} & \textbf{\#sources$=$1} & \textbf{\#sources$=$2} & \textbf{\#sources$=$3} & \textbf{\#sources$=$4} & \textbf{\#sources$=$5} & \textbf{$\Delta$(1$\to$5)} \\
    \midrule
    Alarm Clock       & 0.991 & 0.985 & 0.978 & 0.980 & \textbf{0.969} & $-$0.022 \\
    Toilet Flush      & 1.000 & 0.962 & 0.964 & 0.963 & \textbf{0.954} & $-$0.046 \\
    Music             & 0.995 & 0.968 & 0.966 & 0.955 & \textbf{0.952} & $-$0.043 \\
    Cock A Doodle Doo & 0.995 & 0.982 & 0.969 & 0.957 & \textbf{0.942} & $-$0.053 \\
    Baby Cry          & 0.962 & 0.941 & 0.936 & 0.930 & \textbf{0.918} & $-$0.044 \\
    Cricket           & 0.962 & 0.919 & 0.925 & 0.926 & \textbf{0.916} & $-$0.046 \\
    Siren             & 0.883 & 0.869 & 0.871 & 0.855 & 0.851          & $-$0.032 \\
    Singing           & 0.854 & 0.868 & 0.854 & 0.866 & 0.837          & $-$0.017 \\
    \midrule
    Dog               & 0.964 & 0.907 & 0.905 & 0.885 & 0.886          & $-$0.078 \\
    Ocean             & 0.944 & 0.889 & 0.887 & 0.893 & 0.873          & $-$0.071 \\
    Birds Chirping    & 0.951 & 0.903 & 0.894 & 0.897 & 0.869          & $-$0.082 \\
    Thunderstorm      & 0.925 & 0.883 & 0.876 & 0.865 & 0.856          & $-$0.069 \\
    Car Horn          & 0.898 & 0.880 & 0.851 & 0.842 & 0.821          & $-$0.077 \\
    Computer Typing   & 0.881 & 0.833 & 0.846 & 0.834 & 0.819          & $-$0.062 \\
    Cat               & 0.949 & 0.905 & 0.870 & 0.860 & 0.832          & $-$0.117 \\
    Speech            & 0.905 & 0.826 & 0.819 & 0.803 & 0.799          & $-$0.106 \\
    \midrule
    Glass Breaking    & 0.927 & 0.791 & 0.766 & 0.752 & 0.736          & $-$0.191 \\
    Gunshot           & 0.866 & 0.752 & 0.704 & 0.701 & 0.682          & $-$0.184 \\
    Door Knock        & 0.845 & 0.730 & 0.745 & 0.751 & 0.728          & $-$0.117 \\
    Hammer            & 0.773 & 0.652 & 0.681 & 0.699 & 0.699          & $-$0.074 \\
    \midrule
    \textbf{Macro F1} & \textbf{0.923} & 0.872 & 0.865 & 0.861 & \textbf{0.847} & $-$0.076 \\
    \bottomrule
  \end{tabular}
  }
  \vskip -0.1in
\end{table*}

\subsection{SED Performance Across Durations}

Table~\ref{tab:sed_duration} presents the detailed performance of our fine-tuned AST model across all audio segment duration and source count combinations. Performance generally improves with longer audio segments, as more temporal context allows for better sound event detection. With 10-second segments and a single source, the model achieves 99.4\% accuracy, 91.9\% precision, 95.1\% recall, and 93.5\% F1-score. Notably, even in the most challenging condition (2-second segments with 5 sources), the model maintains 88.1\% accuracy and 74.7\% F1-score.

\begin{table}[!tbp]
  \caption{SED performance across audio lengths and number of target sources. \textmd{Fine-tuned AST model with per-class optimal thresholds.}}
  \label{tab:sed_duration}
  \vskip -0.1in
  \centering
  \setlength{\tabcolsep}{4pt}
  {\footnotesize
  \begin{tabular}{ c c c c c c }
    \toprule
    \textbf{Length} & \textbf{\#Sources} & \textbf{Accuracy} & \textbf{Precision} & \textbf{Recall} & \textbf{F1-Score} \\
    \midrule
    10s & 1 & 0.994 & 0.919 & 0.951 & 0.935 \\
        & 2 & 0.975 & 0.835 & 0.904 & 0.868 \\
        & 3 & 0.964 & 0.843 & 0.911 & 0.876 \\
        & 4 & 0.953 & 0.838 & 0.918 & 0.876 \\
        & 5 & 0.939 & 0.841 & 0.910 & 0.874 \\
    \midrule
    7s  & 1 & 0.994 & 0.930 & 0.940 & 0.935 \\
        & 2 & 0.976 & 0.830 & 0.924 & 0.875 \\
        & 3 & 0.965 & 0.856 & 0.906 & 0.880 \\
        & 4 & 0.952 & 0.842 & 0.912 & 0.876 \\
        & 5 & 0.937 & 0.840 & 0.903 & 0.870 \\
    \midrule
    4s  & 1 & 0.992 & 0.909 & 0.934 & 0.921 \\
        & 2 & 0.971 & 0.796 & 0.901 & 0.845 \\
        & 3 & 0.958 & 0.804 & 0.903 & 0.851 \\
        & 4 & 0.943 & 0.800 & 0.904 & 0.849 \\
        & 5 & 0.923 & 0.797 & 0.884 & 0.838 \\
    \midrule
    3s  & 1 & 0.991 & 0.904 & 0.914 & 0.909 \\
        & 2 & 0.966 & 0.788 & 0.860 & 0.822 \\
        & 3 & 0.949 & 0.771 & 0.871 & 0.818 \\
        & 4 & 0.928 & 0.772 & 0.853 & 0.811 \\
        & 5 & 0.906 & 0.779 & 0.836 & 0.806 \\
    \midrule
    2s  & 1 & 0.989 & 0.884 & 0.888 & 0.886 \\
        & 2 & 0.957 & 0.716 & 0.831 & 0.769 \\
        & 3 & 0.934 & 0.707 & 0.828 & 0.763 \\
        & 4 & 0.909 & 0.716 & 0.806 & 0.758 \\
        & 5 & 0.881 & 0.703 & 0.797 & 0.747 \\
    \bottomrule
  \end{tabular}
  }
  \vskip -0.1in
\end{table}


\section{FUSS External Benchmark Details}
\label{appendix:fuss}

This appendix provides detailed analysis of the FUSS evaluation discussed in \xref{secmain:fuss}.

\subsection{Per-class Results}

Table~\ref{tab:fuss_perclass} compares per-class SI-SDRi between our evaluation set and the FUSS eval set. Of the 20 target classes, 15 are evaluable on FUSS (180 of 1{,}445 foreground sources, 12.5\%). Five classes (alarm clock, baby cry, cock-a-doodle-doo, music, ocean) have zero matchable sources in FUSS due to differences in how sound categories are labeled between FSD50K and our training pipeline. For example, our pipeline defines music as isolated melodies, while FUSS contains instrument-level recordings under a broader category.

\begin{table}[!tbp]
  \caption{Per-class SI-SDRi comparison: FUSS vs.\ our evaluation set. \textmd{15 of 20 classes evaluable. FUSS sources are independently mixed FSD50K recordings.}}
  \label{tab:fuss_perclass}
  \vskip -0.1in
  \centering
  \setlength{\tabcolsep}{4pt}
  {\footnotesize
  \begin{tabular}{ l r r r r }
    \toprule
    \textbf{Class} & \textbf{FUSS $n$} & \textbf{FUSS} & \textbf{Ours} & \textbf{$\Delta$} \\
    \midrule
    car horn        & 4  & +20.94 & 11.70 & +9.24  \\
    door knock      & 15 & +20.62 & 10.10 & +10.52 \\
    dog             & 7  & +17.20 & 10.28 & +6.92  \\
    glass breaking  & 10 & +16.56 & 9.78  & +6.78  \\
    singing         & 21 & +14.95 & 8.64  & +6.31  \\
    cat             & 14 & +12.98 & 9.94  & +3.04  \\
    birds chirping  & 9  & +10.04 & 10.72 & $-$0.68 \\
    toilet flush    & 6  & +9.49  & 8.81  & +0.68  \\
    speech          & 34 & +7.08  & 10.10 & $-$3.02 \\
    hammer          & 9  & +5.64  & 10.33 & $-$4.69 \\
    gunshot         & 35 & +5.32  & 8.85  & $-$3.53 \\
    thunderstorm    & 1  & +2.97  & 10.05 & $-$7.09 \\
    computer typing & 8  & +0.58  & 8.19  & $-$7.61 \\
    cricket         & 3  & $-$11.13 & 10.90 & $-$22.03 \\
    siren           & 4  & $-$13.12 & 9.30 & $-$22.43 \\
    \bottomrule
  \end{tabular}
  }
  \vskip -0.1in
\end{table}

Classes where FUSS exceeds our evaluation set (car horn, door knock, dog, glass breaking) benefit from FUSS's clean, single-source FSD50K recordings. Classes with very few samples (cricket: $n{=}3$, siren: $n{=}4$, thunderstorm: $n{=}1$) exhibit high variance and are statistically unreliable.

\subsection{Low-Performance Cause Analysis}

The low-performing classes on FUSS are not indicative of model deficiency but rather reflect fundamental differences between the FUSS and our evaluation set data pipelines:

\squishlist
    \item \textbf{Semantic background sources.} Our model was trained with environmental noise backgrounds (WHAM!, UrbanSound8K), which are non-semantic and never overlap with our 20 target classes. FUSS instead uses a single FSD50K recording as background, which can belong to one of our target classes (e.g., door knock, singing, cat). This creates ambiguity between ``what to extract'' and ``what to ignore'' that the model has never encountered during training.

    \item \textbf{Extreme sample imbalance.} Our evaluation set contains approximately 300 samples per class, whereas FUSS coverage ranges from 1 to 35 per class, with 7 classes having fewer than 10 samples.
\squishend

\subsection{FUSS Label Distribution}

\begin{figure*}[t!]
\centering
\includegraphics[width=0.9\linewidth]{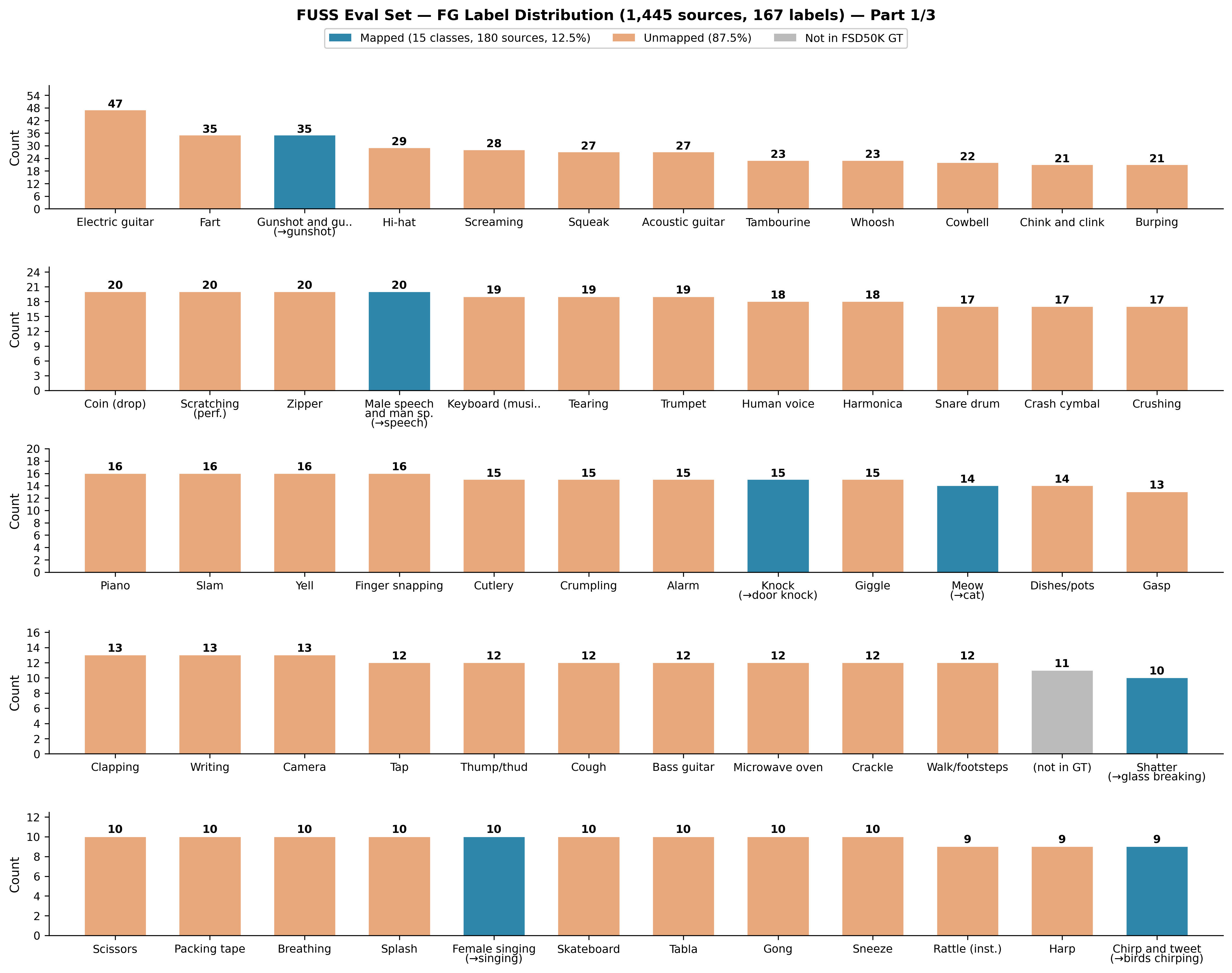}
\vskip -0.1in
\caption{Complete FG source label distribution of the FUSS eval set --- Part 1/3 (top-ranked labels). \textmd{1{,}445 sources across 167 unique labels. Blue: mapped to our 20 target classes (180 sources, 12.5\%). Orange: unmapped FSD50K categories. Gray: not in FSD50K ground truth. Continued in Figs.~\ref{fig:fuss_label_dist_2} and \ref{fig:fuss_label_dist_3}.}}
\label{fig:fuss_label_dist_1}
\vskip -0.15in
\end{figure*}
\begin{figure*}[t!]
\centering
\includegraphics[width=0.9\linewidth]{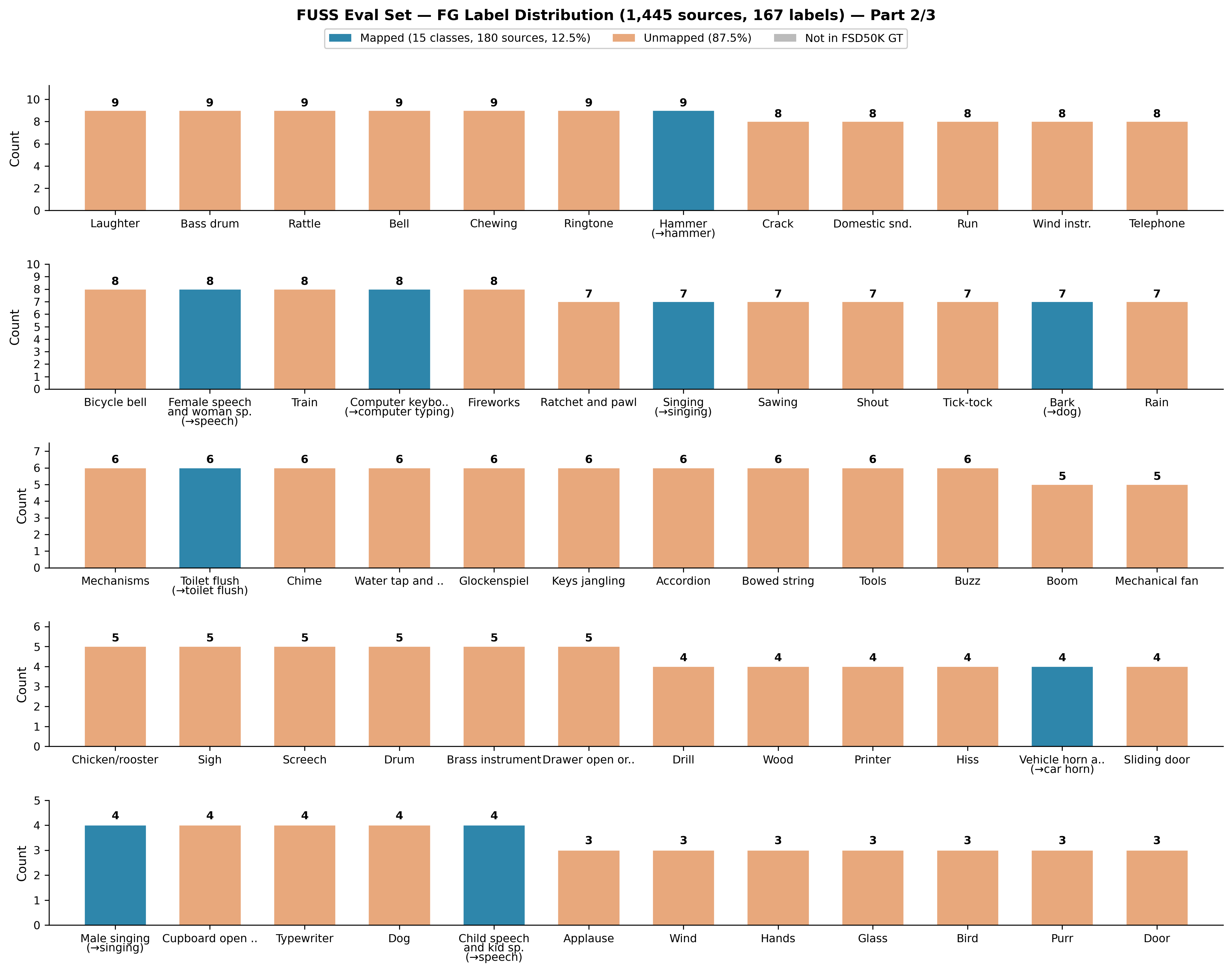}
\vskip -0.1in
\caption{Complete FG source label distribution of the FUSS eval set --- Part 2/3 (mid-ranked labels). \textmd{Continued from Fig.~\ref{fig:fuss_label_dist_1}.}}
\label{fig:fuss_label_dist_2}
\vskip -0.15in
\end{figure*}
\begin{figure*}[t!]
\centering
\includegraphics[width=0.9\linewidth]{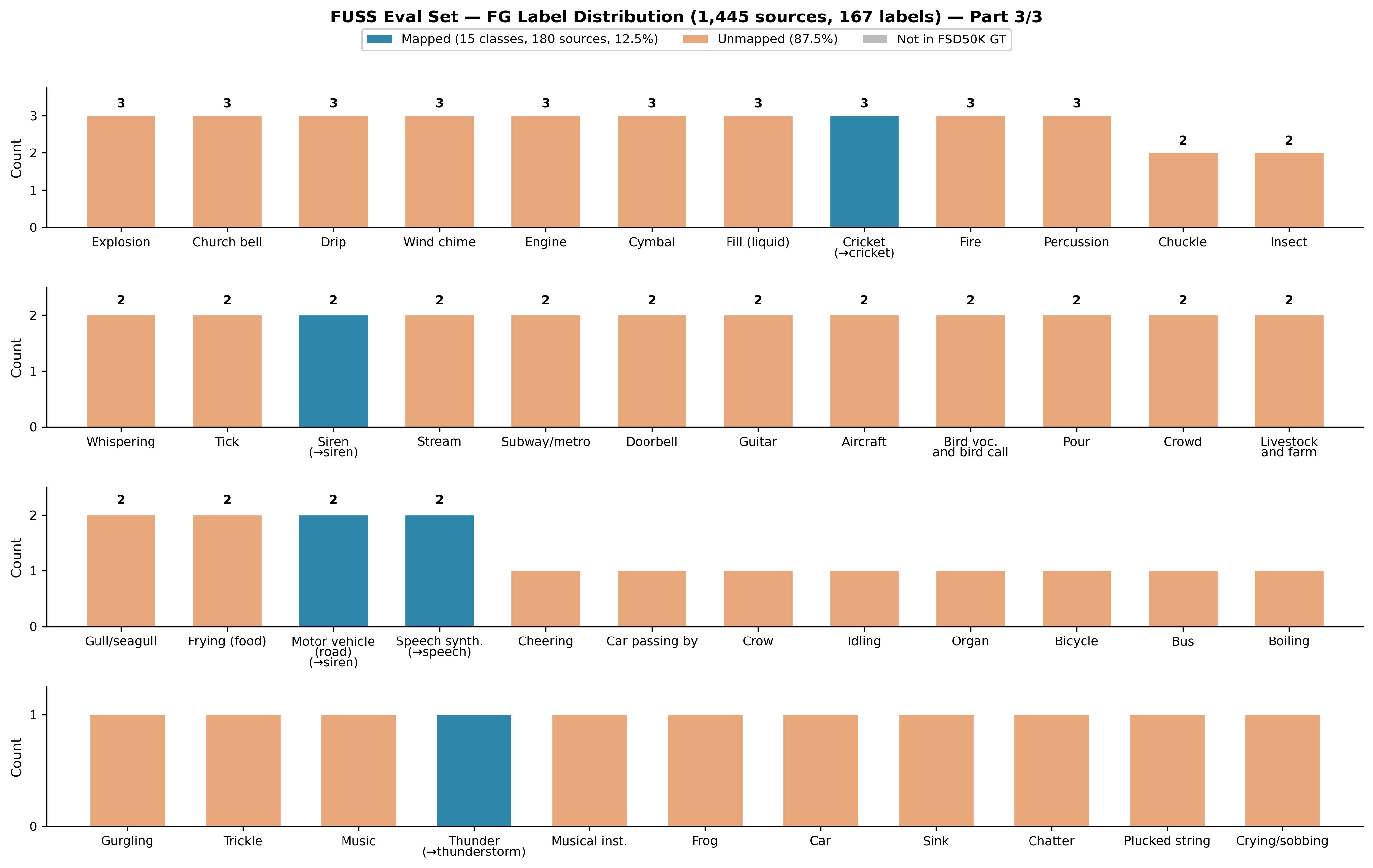}
\vskip -0.1in
\caption{Complete FG source label distribution of the FUSS eval set --- Part 3/3 (lowest-ranked labels). \textmd{Continued from Fig.~\ref{fig:fuss_label_dist_2}. Most labels in this range have count $\leq$ 5, limiting statistical reliability of per-class analysis on FUSS.}}
\label{fig:fuss_label_dist_3}
\vskip -0.15in
\end{figure*}

Figs.~\ref{fig:fuss_label_dist_1}--\ref{fig:fuss_label_dist_3} show the complete label distribution of all 1{,}445 foreground sources in the FUSS eval set. Only 180 sources (12.5\%) across 15 classes map to our 20-class vocabulary. The unmapped 87.5\% are predominantly musical instruments (electric guitar, hi-hat, acoustic guitar) and domestic/body sounds (fart, burping, coin dropping) that fall outside our target class definitions. This coverage imbalance limits per-class reliability on FUSS. The overall SI-SDRi (+9.93\,dB on FUSS vs.\ +10.15\,dB on ours) is a more meaningful metric.

\clearpage

\clearpage
\section{Artifact Appendix}

\subsection{Abstract}

This artifact includes source code, pre-trained checkpoints, dataset preparation scripts, and evaluation pipelines for the TSE and SED models presented in the paper.
Pre-trained checkpoints on HuggingFace enable evaluation without retraining.
A mini dataset (${\sim}$250\,MB) is provided for quick pipeline verification (${\sim}$5\,min on GPU); the full dataset (${\sim}$130\,GB) reproduces all paper results.

\subsection{Artifact check-list (meta-information)}

{\small
\begin{itemize}[itemsep=1pt,parsep=2pt,topsep=3pt,partopsep=0pt,leftmargin=0em,itemindent=1em,labelwidth=1em,labelsep=0.5em]
  \item {\bf Algorithm:} Dual-path time-frequency model with FiLM conditioning (TSE); fine-tuned AST (SED)
  \item {\bf Program:} Python / PyTorch
  \item {\bf Compilation:} N/A (interpreted)
  \item {\bf Model:} 11 TSE checkpoints + Waveformer baseline + 1 SED model; hosted on HuggingFace
  \item {\bf Data set:} FSD50K, ESC-50, MUSDB18, DISCO, TAU-2019, CIPIC HRTF (${\sim}$130\,GB total); mini subset (${\sim}$250\,MB)
  \item {\bf Run-time environment:} Linux, Python~3.11, PyTorch~2.0+, CUDA~12.x; Docker image provided
  \item {\bf Hardware:} Any CUDA-compatible NVIDIA GPU (tested on A40)
  \item {\bf Run-time state:} No special requirements
  \item {\bf Execution:} Single GPU, no process pinning required
  \item {\bf Metrics:} SNRi, SI-SNRi (dB) for TSE; accuracy, precision, recall, F1-score for SED
  \item {\bf Output:} Console + CSV/JSON files with per-class and aggregate metrics
  \item {\bf Experiments:} Shell scripts + Docker
  \item {\bf Disk space:} ${\sim}$130\,GB (full) or ${\sim}$500\,MB (mini + models + code)
  \item {\bf Preparation time:} ${\sim}$5\,min (mini) or ${\sim}$1--2\,h (full dataset)
  \item {\bf Experiment time:} ${\sim}$5\,min (mini), ${\sim}$10\,h (full), ${\sim}$24\,h (training)
  \item {\bf Publicly available?:} Yes (GitHub + HuggingFace)
  \item {\bf Code licenses:} MIT
  \item {\bf Data licenses:} Mixed per dataset (CC-BY, CC-BY-NC, academic)
  \item {\bf Workflow framework:} Shell scripts, Docker
  \item {\bf Archived:} GitHub + HuggingFace
\end{itemize}
}

\subsection{Description}

\vskip 0.03in\noindent{\bf How to access.}\\
{\bf Code:} {\small\url{https://github.com/ooshyun/fine_grained_soundscape_control}}\\
{\bf TSE models:} {\small\url{https://huggingface.co/ooshyun/fine_grained_soundscape_control}}\\
{\bf SED model:} {\small\url{https://huggingface.co/ooshyun/sound_event_detection}}\\
{\bf Datasets:} {\small\url{https://huggingface.co/datasets/ooshyun/fine-grained-soundscape}}

\vskip 0.03in\noindent{\bf Hardware dependencies.}
Any CUDA-compatible NVIDIA GPU (tested on A40, 48\,GB).

\vskip 0.03in\noindent{\bf Software dependencies.}
Python~3.11, PyTorch~2.0+, torchaudio, Lightning~2.0+, Transformers~4.30+ ($<$5.0). See \texttt{requirements.txt}.

\vskip 0.03in\noindent{\bf Datasets.}
The evaluation uses six public datasets totaling ${\sim}$130\,GB.

\vskip 0.03in\noindent{\bf Option~A:} Download the public tar from
\url{https://semantichearing.cs.washington.edu/BinauralCuratedDataset.tar},
or run {\small\texttt{bash scripts/setup\_dataset.sh -{}-output\_dir /path/to/data}}.

\vskip 0.03in\noindent{\bf Option~B (recommended for quick verification):} A pre-built mini dataset (${\sim}$250\,MB, 1~file per class) is available on HuggingFace.

\vskip 0.03in\noindent{\bf Models.}
Pre-trained models are downloaded automatically by the evaluation scripts.
11~TSE checkpoints (3~architectures $\times$ 3~FiLM variants + 2~multi-output + Waveformer) and one fine-tuned AST for SED.

\subsection{Installation}

\noindent{\bf Option A Local:}
{\small\texttt{git clone -{}-recursive <repo\_url>}}, then
{\small\texttt{pip install -r requirements.txt}} in a Python~3.11 venv.

\vskip 0.03in\noindent{\bf Option B Docker (recommended):}\par
{\small\texttt{bash docker/build.sh}}.
Use {\small\texttt{-{}-shm-size=2g}} when running (required for SED evaluation).

\subsection{Experiment workflow}

\vskip 0.03in
\noindent\resizebox{\columnwidth}{!}{%
\begin{tabular}{@{}llll@{}}
{\bf Mode} & {\bf Interface} & {\bf Time} & {\bf Coverage} \\
\hline
Mini & Shell (one command) & ${\sim}$5\,min & Table~\ref{tab:orangepi_eval} (2 models); Fig.~\ref{fig:sed_comparison} (subset) \\
Full & Shell / Docker & ${\sim}$10\,h & Tables~\ref{tab:orangepi_eval}--\ref{tab:latency}; Fig.~\ref{fig:sed_comparison} (complete) \\
\end{tabular}
}

\vskip 0.03in\noindent{\bf (1) Mini evaluation} (recommended first step):

\noindent Downloads mini dataset, extracts, and runs TSE~+~SED evaluation (50 samples each).\par
{\small\texttt{bash scripts/eval/eval\_mini.sh [data\_dir] [output\_dir]}}

\vskip 0.03in\noindent{\bf (2) Full evaluation} (requires full dataset):

\vskip 0.03in
\noindent\resizebox{\columnwidth}{!}{%
\begin{tabular}{@{}ll@{}}
  Table~\ref{tab:orangepi_eval} (TSE models) & \texttt{scripts/eval/run\_tse.sh} \\
  Table~\ref{tab:orangepi_eval} (multi-output) & \texttt{scripts/eval/run\_multiout.sh} \\
  Table~\ref{tab:ablation_film} (FiLM ablation) & \texttt{scripts/eval/run\_ablation.sh} \\
  Fig.~\ref{fig:sed_comparison} (SED) & \texttt{scripts/eval/run\_sed.sh} \\
  All (Docker) & \texttt{bash docker/eval\_all.sh /path/to/data} \\
\end{tabular}
}

\subsection{Evaluation and expected results}

Table~\ref{tab:orangepi_eval} compares four TSE architectures on 20 target classes with a single source and one output channel. OrangePi achieves 11.99\,dB SNRi and 11.27\,dB SI-SNRi, while the Waveformer baseline reaches 7.29 and 5.58\,dB respectively.
Table~\ref{tab:ablation_film} presents the FiLM ablation with 5 output channels and 1--5 targets in the mixture; applying FiLM to all blocks (12.26\,dB SNRi) outperforms first-block-only placement (11.76\,dB).
Fig.~\ref{fig:sed_comparison} compares SED models (fine-tuned AST, YAMNet, baseline AST) across 1--5 concurrent sources on 5-second segments; the fine-tuned AST maintains high accuracy even with 5 sources.
Table~\ref{tab:latency} reports the accuracy-latency trade-off across model sizes and window configurations; reproducing this table requires full training and is not covered by the evaluation-only scripts.

Acceptable variation is $\pm$1.0\,dB for SNRi/SI-SNRi and $\pm$2\% for classification metrics, as evaluation uses on-the-fly binaural mixture synthesis with randomized spatial positions.
The in-the-wild evaluation and dynamic interface study involve human subjects and cannot be reproduced computationally.

\end{document}